# Efficient GPU-Accelerated Training of a Neuroevolution Potential with Analytical Gradients

Hongfu Huang,[a] Junhao Peng,[b] Kaiqi Li,[c] Jian Zhou,[d] and Zhimei Sun[*d]

[a] School of Integrated Circuit Science and Engineering, Beihang University, Beijing, 100191, China.

[b] Guangdong Provincial Key Laboratory of Sensing Physics and System Integration Applications, School of Physics and Optoelectronic Engineering, Guangdong University of Technology, Guangzhou, Guangdong 510006, China.

[c] National Key Laboratory of Spintronics, Hangzhou International Innovation Institute, Beihang University, Hangzhou 311115, China.

[d] School of Materials Science and Engineering, Beihang University, Beijing, 100191, China.

[*] Corresponding author: zmsun@buaa.edu.cn

## ABSTRACT

Machine-learning interatomic potentials (MLIPs) such as neuroevolution potentials (NEP) combine quantum-mechanical accuracy with computational efficiency significantly accelerate atomistic dynamic simulations. Trained by derivative-free optimization, the normal NEP achieves good accuracy, but suffers from inefficiency due to the high-dimensional parameter search. To overcome this problem, we present a gradient-optimized NEP (GNEP) training framework employing explicit analytical gradients and the Adam optimizer. This approach greatly improves training efficiency and convergence speedily while maintaining accuracy and physical interpretability. By applying GNEP to the training of Sb-Te material systems (datasets include crystalline, liquid, and disordered phases), the fitting time has been substantially reduced —often by orders of magnitude—compared to the NEP training framework. The fitted potentials are validated by DFT reference calculations, demonstrating satisfactory agreement in equation of state and radial distribution functions. These results confirm that GNEP retains high predictive accuracy and transferability while considerably improved computational efficiency, making it well-suited for large-scale molecular dynamics simulations.

## I. INTRODUCTION

Machine-learning interatomic potentials (MLIPs) have emerged as powerful tools to bridge the accuracy of first-principles methods and the efficiency of classical force fields in materials simulations[1–5]. By learning potential energy surfaces from quantum-mechanical data, MLIPs enable large-scale molecular dynamics (MD) simulations with near-ab initio accuracy, thereby facilitating investigations of complex phenomena such as phase transitions, diffusion, transport and mechanical properties[6–12]. Early examples include High-dimensional neural network potentials[13], Moment tensor potential[14], Spectral neighbor analysis potential[15] and Gaussian approximation potentials[16], which demonstrated that neural networks and other regressors can predict atomic energies and forces with high fidelity to ab initio calculations. More recently, the neuroevolution potential (NEP) framework was introduced which employs a feed-forward neural network to model atomic energies as functions of local environment descriptors constructed from orthogonal polynomial bases (Chebyshev and Legendre) [17]. A key distinction of NEP is its training procedure: it originally relies on the separable natural evolution strategy (SNES) – a derivative-free optimization algorithm – to fit the network parameters to reference data[18]. This approach circumvented the need for back-propagating gradients through the complex descriptor and network structure, simplifying implementation and avoiding external machine-learning libraries. Indeed, the NEP model was implemented fully on graphics processing units (GPUs) within the GPUMD package[19], achieving high computational speed for MD simulations (on the order of millions of atom-steps per second on modern hardware).

Despite its success in accuracy and parallel efficiency, the SNES-based training strategy presents a significant computational bottleneck. Derivative-free evolutionary optimization requires evaluating a large population of trial parameter sets or numerous random perturbations to approximate gradients, which becomes increasingly costly as the number of model parameters grows[18,20]. In the NEP framework, the descriptor and neural network can contain tens of thousands of parameters, making the search space high-dimensional. As a result, SNES training can demand many iterations to converge, consuming substantial compute time even with GPU acceleration. In practice, while SNES's global search capability can be robust, it often converges slowly compared to gradient-based methods for large neural networks. This slower convergence delays the development cycle of new potentials and limits the efficient exploration of hyperparameters or active-learning iterations.

Therefore, there is strong motivation to incorporate analytical gradients into the NEP training process to enable conventional backpropagation and fast gradient-based optimization (e.g. Adam[21,22]). Beyond computational speed, one can inspect the gradient components to see, for example, which descriptor term is contributing most to a certain force component, potentially linking that to a physical interaction. This transparency is valuable for understanding the material's physics and for explaining why the potential makes certain predictions. Additionally, ensuring energy-force consistency by analytical differentiation means our model strictly obeys energy conservation, which is crucial for correct simulation of dynamical properties like thermal conductivity.

Despite the advantages, several challenges associated with the current approach must be acknowledged. Neural network training in high-dimensional parameter spaces is well known to encounter many local minima or flat regions[23,24]. Fortunately, our tests did not reveal any convergence issues. However, it is possible that for certain systems or if the model is made more complex, the optimizer could get trapped in a suboptimal minimum. In contrast, SNES, as a global search method, possesses a theoretical advantage in finding the global optimum given sufficient iterations, due to its inherent ability to escape local minima through random exploration[18,25]. Thus, one potential risk of relying solely on gradient-based training is that we might settle for a non-global optimum of the loss function, which could correspond to, e.g., slightly worse accuracy or transferability. In practice, this risk can be mitigated by performing multiple training runs with different random initializations and selecting the best-performing model, or by employing active learning strategies to enhance the diversity of the training dataset[2,19,26,27].

In this work, we present an enhanced NEP training approach that leverages explicit analytical gradients of the potential with respect to its parameters, implemented on GPUs via CUDA. Our goal is to dramatically accelerate the optimization of NEP models by replacing the SNES optimizer with gradient-based algorithms, while preserving the accuracy and physical interpretability of the model. We derive the analytical expressions for the gradient of the loss function (based on energy, force and virial errors) with respect to both network weights and descriptor parameters. These gradients are computed efficiently on the GPU, enabling the use of the Adam optimizer to speed up convergence. By eliminating the costly evolutionary loop, our approach significantly reduces training time and computational resource usage. Gradient-optimized Neuroevolution potential (GNEP) training approach was utilized to develop the interatomic potential for the Sb-Te system. The fitted potential was validated by computing the equation of state (EOS) for the target material and conducting large-scale molecular dynamics simulations. To assess the transferability of our interatomic potential, we analyzed radial distribution functions (RDFs) from the simulations and compared them against reference data obtained from DFT calculations.

## II. THEORY

This section presents the theoretical foundation of our MLIP based on the NEP.

A. Energy, Forces and Virial stress

The NEP represents the total potential energy of a system as a sum of environment-dependent atomic energies[28]:

$$E = \sum_i E_i = \sum_i \mathcal{F}(\boldsymbol{G_i}) \in \mathbb{R}^1 \qquad (1)$$

Here, $E_i$ is the energy of atom $i$, computed from a descriptor vector $\boldsymbol{G_i}$ via a neural network function $\mathcal{F}(\cdot)$. For each atom $i$, a local descriptor vector $\boldsymbol{G_i}$ is constructed to encode the geometric arrangement of its neighboring atoms $\mathcal{N}^i$ within a cutoff radius $r_c$:

$$\mathcal{N}^i = \{j \neq i \mid \|\boldsymbol{r_{ij}}\| < r_c\}, \ \boldsymbol{r_{ij}} \equiv \boldsymbol{r_j} - \boldsymbol{r_i} \qquad (2)$$

This ensures the model is both efficient and physically local[28]. Descriptors constructed on this basis are designed to preserve translational, rotational, and permutation invariance.

Forces are obtained via the negative gradient of the energy with respect to atomic positions:

$$F_i = -\nabla_{r_{ij}} E = \sum_{j \in \mathcal{N}^i} F_{ij} = -\sum_{j \in \mathcal{N}^i} F_{ji} = \sum_{j \in \mathcal{N}^i} \left( \frac{\partial E_i}{\partial r_{ij}} - \frac{\partial E_j}{\partial r_{ji}} \right) \in \mathbb{R}^3 \qquad (3)$$

Here, $F_i$ represents the total force acting on atom $i$, which is the vector sum of the interaction forces $F_{ij}$ exerted on atom $i$ by all other atoms ($j \neq i$). This expression ensures Newton's third law satisfied in its weak form ($F_{ij} = -F_{ji}$), thereby preserving the total momentum of the system[29]. The analytical derivations for $\frac{\partial E_i}{\partial r_{ij}}$ includes the derivatives of energy with respect to $x_{ij}, y_{ij}, z_{ij}$ directions. These are calculated using the chain rule and backpropagation in neural networks.

The per-atom virial stress is defined to characterize internal mechanical stress in molecular dynamics simulations:

$$V_i = \sum_{j \in \mathcal{N}^i} r_{ij} \otimes \frac{\partial E_j}{\partial r_{ji}} \in \mathbb{R}^6 \qquad (4)$$

Where ⊗ denotes the outer product of vectors. The total stress of the system can then be assembled from the per-atom virials.

B. Descriptors vector construction

These descriptors consist of both radial terms (two-body) and angular terms (many-body up to four- or five-body) that are carefully designed to be permutation-invariant and differentiable.

The radial part is typically built by expanding each neighbor distance $r_{ij}$ (for neighbor $j$ of atom $i$) in a set of orthogonal basis functions (Chebyshev polynomials here) multiplied by a smooth cutoff function:

$$\boldsymbol{q_n} = \{q_n^i\} \in \mathbb{R}^N = \sum_{j \in \mathcal{N}^i} g_n(r_{ij}) \tag{5}$$

$$g_n(r_{ij}) = \sum_{k=0}^{N_{\text{bas}}^R} c_{nk}^{IJ} f_k(r_{ij}) = \sum_{k=0}^{N_{\text{bas}}^R} c_{nk}^{IJ} \frac{T_k(\xi) + 1}{2} f_c(r_{ij}) \tag{6}$$

$c_{nk}^{IJ}$ are trainable expansion coefficients, where $I$ and $J$ represent the element types of atoms $i$ and $j$, respectively. The set of basis functions $f_k(r_{ij})$ are indexed by $k = 0, 1, \ldots, N_{\text{bas}}^R$.

Angular descriptors in NEP capture many-body correlations, including three-body, four-body, and five-body angular interactions. These components are essential for accurately modeling local bonding geometry and are constructed based on spherical harmonics. The detailed derivation of the angular descriptors, especially the four- and five-body terms, can be found in Ref.[9]. This work focuses on the three-body terms, using the addition theorem of spherical harmonics to express directional dependence:

$$\boldsymbol{q_{nl}} = \{q_{nl}^i\} \in \mathbb{R}^N = \sum_{m=0}^{2l} C_{lm} (S_{nlm}^i)^2 \tag{7}$$

$$\boldsymbol{S_{nlm}} = \{S_{nlm}^i\} \in \mathbb{R}^N = \sum_{j \neq i} \frac{g_n(r_{ij})}{r_{ij}^l} b_{lm}(x_{ij}, y_{ij}, z_{ij}) \tag{8}$$

Here, $l = 1, 2, \ldots, 8$ and $m$ represent the degree and order of the Legendre polynomial. $b_{lm}(x_{ij}, y_{ij}, z_{ij})$ are real spherical harmonics expressed in Cartesian form, and $C_{lm}$ are constant coefficients.

Detailed expressions for $T_k(\xi)$, the cutoff function $f_c(r_{ij})$, chain rule and backpropagation in neural networks are provided in **Appendix A**. Expressed real spherical harmonics in Cartesian for are provided int **Appendix B**. Gradients of these descriptors with respect to atomic positions and model parameters are listed in **Appendix C**.

C. Loss Function and Gradients

Each atomic descriptor $\boldsymbol{G}_i = \{q_n^i, q_{nl}^i\}$ is fed into a neural network $\mathcal{F}(\cdot)$ to predict the atomic energy $E_i$. The model parameters—including descriptor coefficients $c_{nk}^{IJ}$ and network weights/bias—are optimized by minimizing a weighted loss function:

$$\mathcal{L} = \lambda_e \mathcal{L}_e + \lambda_f \mathcal{L}_f + \lambda_v \mathcal{L}_v \tag{9}$$

$$\mathcal{L}_e = \frac{1}{N_{str}} \sum_{n=1}^{N_{str}} \left( E_i^{pre}(n) - E_i^{tar}(n) \right)^2 \tag{10}$$

$$\mathcal{L}_f = \frac{1}{3N} \sum_{i,\alpha} \left( F_{i,\alpha}^{pre} - F_{i,\alpha}^{tar} \right)^2 \tag{11}$$

$$\mathcal{L}_v = \frac{1}{6N_{str}} \sum_{n=1}^{N_{str}} \sum_{l=1}^{6} \left( V_{i,l}^{pre}(n) - V_{i,l}^{tar}(n) \right)^2 \tag{12}$$

Here, $\lambda_e$, $\lambda_f$ and $\lambda_v$ are the weights for energies, forces, and virials, respectively, which control their relative importance during training. $N_{str}$ is the total number of structures in the batch. $E_i^{pre}(n)$ and $E_i^{tar}(n)$ are the predicted and reference (per-atom) energies for the $n-th$ structure. $N$ is the total number of atoms in the batch. $F_{i,\alpha}^{pre}$ and $F_{i,\alpha}^{tar}$ represent the predicted and target forces for the $i-th$ atom in the current batch, with $\alpha \in \{x, y, z\}$ indicating the Cartesian direction. $V_{i,l}^{pre}(n)$ and $V_{i,l}^{tar}(n)$ denote the predicted and reference values of the $l-th$ virial tensor component (per-atom virial) for the $n-th$ structure.

According to the above formula, the gradient used to update the parameter ($c_{nk}^{IJ}$) is obtained as:

$$\frac{\partial \mathcal{L}}{\partial c_{nk}^{IJ}} = \lambda_e \frac{\partial \mathcal{L}_e}{\partial c_{nk}^{IJ}} + \lambda_f \frac{\partial \mathcal{L}_f}{\partial c_{nk}^{IJ}} + \lambda_v \frac{\partial \mathcal{L}_v}{\partial c_{nk}^{IJ}} \tag{13}$$

$$\frac{\partial \mathcal{L}_e}{\partial c_{nk}^{IJ}} = \frac{2}{N_{str}} \sum_{n=0}^{N_{str}} \left( E_i^{pre}(n) - E_i^{tar}(n) \right) \times \frac{\partial E_i}{\partial c_{nk}^{IJ}} \quad (14)$$

$$\frac{\partial \mathcal{L}_f}{\partial c_{nk}^{IJ}} = \frac{2}{3N} \sum_{i,\alpha} \left( F_{i,\alpha}^{pre} - F_{i,\alpha}^{tar} \right) \times \frac{\partial F_i}{\partial c_{nk}^{IJ}} \quad (15)$$

$$\frac{\partial \mathcal{L}_v}{\partial c_{nk}^{IJ}} = \frac{2}{6N_{str}} \sum_{n=0}^{N_{str}} \sum_{l=0}^{6} \left( V_{i,l}^{pre}(n) - V_{i,l}^{tar}(n) \right) \times \frac{\partial V_i}{\partial c_{nk}^{IJ}} \quad (16)$$

D. Learning rate schedules

To improve convergence and stabilize training, we adopt a cosine decay learning rate (LR) schedule with an initial warmup phase[30,31]. This schedule is smooth and continuous, allowing the learning rate to increase gradually at the beginning of training before decaying in a cosine manner. The learning rate $lr$ at a given optimization step $s$ is updated dynamically based on the overall training progress.

Specifically, the learning rate increases linearly from the minimum $lr_{stop}$ to the initial maximum $lr_{start}$ during a warmup phase spanning warmup $N_{warmup}$ steps. After warmup, the learning rate decays according to a cosine schedule. Formally, at global step $s$, the learning rate $lr(s)$ is defined as:

$$lr(s) = \begin{cases} lr_{stop} + \frac{s}{N_{warmup}} (lr_{start} - lr_{stop}), & 0 \leq s < N_{warmup} \\ lr_{stop} + (lr_{start} - lr_{stop}) \times \frac{1}{2} \left( 1 + \cos\left( \pi \frac{s - N_{warmup}}{N_{total} - N_{warmup}} \right) \right), & N_{warmup} \leq s \leq N_{total} \end{cases} \quad (17)$$

Where $N_{warmup} = N_{warmup\_epochs} \times N_{batches\_per\_epoch}$, and $N_{total} = N_{epochs} \times N_{batches\_per\_epoch}$. Here, $lr_{start}$ is the initial (maximum) learning rate, and $lr_{stop}$ is the minimum learning rate.

E. DFT calculation and AIMD simulations

The density functional theory (DFT) calculations were conducted using the Vienna Ab initio Simulation Package (VASP) with projector augmented wave (PAW) method. The Perdew-Burke-Ernzerhof (PBE) functional within the generalized gradient approximation (GGA) framework was employed for exchange-correlation interactions. Self-consistent cycles were performed with a plane-wave cutoff energy of 600 eV, maintaining a convergence criterion of 1 x 10$^{-8}$ eV. The k-point mesh for Brillouin zone sampling was generated using VASPKIT with a density of $2\pi \times 0.02$ Å$^{-1}$ in reciprocal space. Van der Waals interactions were incorporated via the DFT-D3 method with zero damping.

For the ab initio molecular dynamics (AIMD) simulations used in comparison, we employed the CP2K software package, which implements a mixed Gaussian and plane-wave (GPW) approach. All calculations utilized Goedecker-Teter-Hutter (GTH) pseudopotentials under periodic boundary conditions. A cutoff energy of 350 Rydberg was used and Brillouin zone sampling was performed at the $\Gamma$-point only. The AIMD trajectories were generated within the canonical NVT ensemble using the canonical sampling through velocity rescaling (CSVR) thermostat, with a 1 fs integration timestep throughout all simulations.

## III. RESULTS

A. Model training

The Sb-Te phases include a wide range of configurations: disordered structures, including liquid and amorphous phases, and ordered structures, containing metastable and stable crystal structure. The components involved contain elementary Sb and Te, binary compounds of $Sb_2Te$, $SbTe$, and $Sb_2Te_3$. Moreover, point defective structures, dimer structures and surface structures are added into the dataset too. A total of 1527 configurations were used for training and 223 for testing.

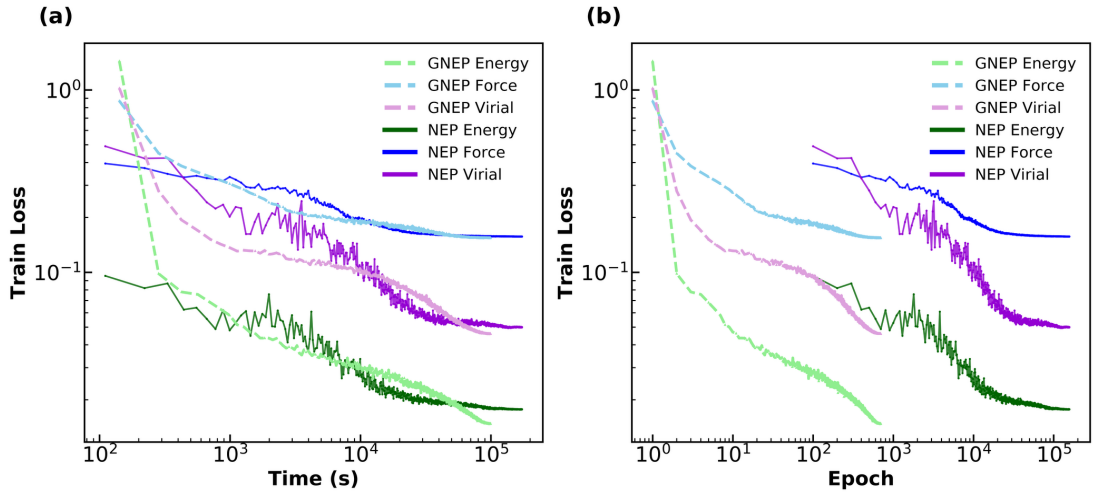

**Figure 1** Training RMSE for energy, force and virial predictions on the Sb-Te dataset, comparing NEP (solid lines) and GNEP (dash–dot lines). (a) Training loss vs. wall-clock time. (b) Training loss vs. the number of epochs. Unless otherwise specified, all GNEP models are trained with mini-batches of size 1, while NEP models use full-batch updates.

We compared the performance of our gradient-based training to that of the original SNES-based approach. As shown in **Figure 1a**, the energy loss under both methods exhibits similar convergence trends; however, the GNEP curve is smoother and ultimately reaches a lower final value. For example, the force loss (blue curve) decreases more steeply in the early training phase, reaching a comparable final value while requiring significantly less time compared to NEP. **Figure 1b** shows the same training curves as a function of epoch. The GNEP reaches low-error plateaus within tens to hundreds of epochs, while NEP requires thousands. For example, the force RMSE falls below 200 meV/Å within ~50 epochs using GNEP, compared to several thousand epochs for NEP to reach similar accuracy.

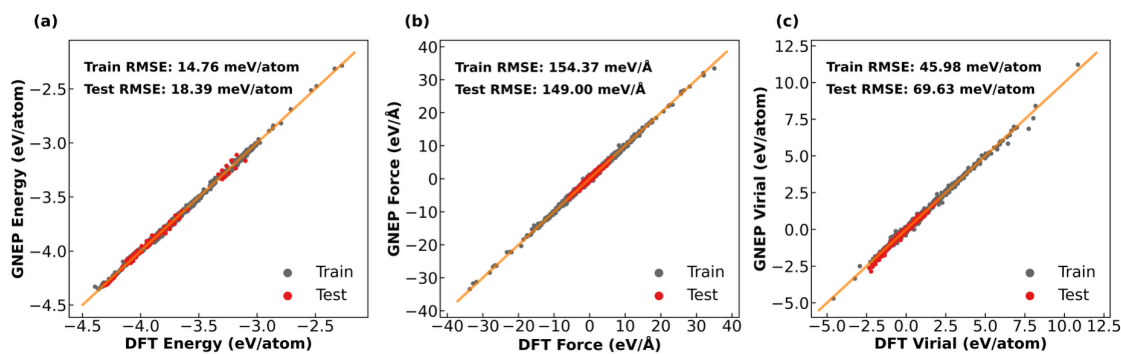

**Figure 2** Correlation between predicted and DFT-computed energies, forces, and virials for the NEP Model. (a-c) Results for the training and test datasets for energy, force, and virial predictions, respectively.

Prediction accuracy is validated in **Figure 2**, showing close agreement with DFT values. The result shows a training root mean square error (RMSE) of 14.76 meV/atom and a test RMSE of 18.39 meV/atom for energy, 154.37 and 149.0 meV/Å for force, and 45.98 and 69.63 meV/atom for virial stress components. The small errors and consistency between training and test sets indicate excellent generalization capability.

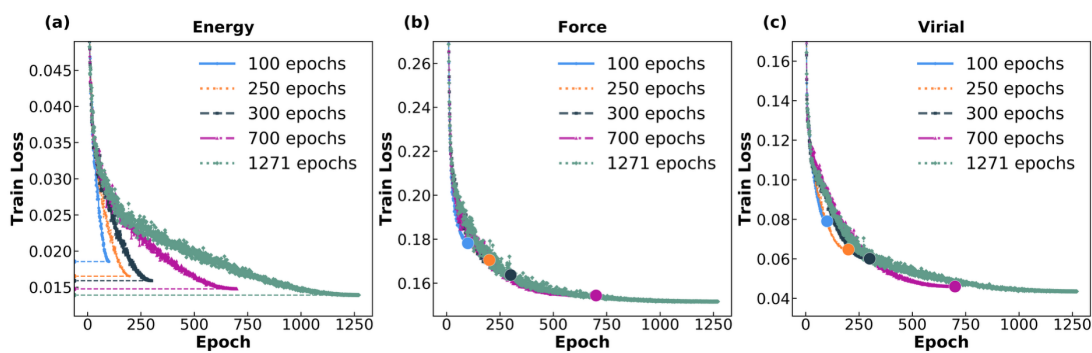

**Figure 3** Training loss curves for (a) energy, (b) force, and (c) virial predictions on the Sb–Te dataset using 100, 200, 300, 700, and 1271 training epochs, respectively.

We also investigated the effect of training duration and batch size on model performance. As shown in **Figure 3**, extending the number of training epochs from 100 to 1271 (specifically, 100, 200, 300, 700, and 1271 epochs) led to only marginal improvements across all three loss components (energy, force, and virial), illustrating the diminishing returns of prolonged training. Excessively long training increases computational cost while yielding limited gains, likely due to the learning rate decaying too slowly. As a result, the optimizer may continue to explore within a narrow region at relatively high learning rates, making it difficult to converge precisely to deeper local minima.

**Figure 4** further shows that smaller batch sizes (e.g., batch size = 1) result in lower final total loss after 100 epochs but require significantly more wall-clock time. In contrast, larger batch sizes reduce training time but converge to higher loss values, reflecting a trade-off between computational efficiency and model precision.

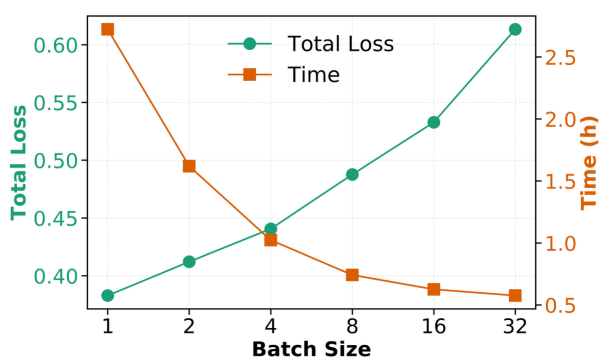

**Figure 4** Final total loss and total training time for different batch sizes after 100 epochs.

Overall, gradient-based training improved stability and produced a smoother, more monotonic reduction in training error. This stability helps in knowing when to stop training (we can use early stopping once the validation error stagnates). In addition to faster convergence, the training typically achieves a final loss comparable to (or at least not worse than) that obtained using SNES in practice. For the systems we tested, this indicates no observable degradation in final accuracy (see **Supporting Information** for results on additional systems).

B. DFT Validation

To evaluate the accuracy of the GNEP model in describing static properties, we first computed the EOS curves of two representative crystalline Sb–Te compounds: $Sb_2Te$ in the $P\bar{3}m1$ structure and $Sb_2Te_3$ in the $R\bar{3}m$ structure. As shown in **Figure 5**, the GNEP predictions closely follow the DFT results, accurately capturing the equation-of-state characteristics. This strong agreement demonstrates the reliability and precision of the model in structural prediction.

We further assessed the model's ability to describe dynamic structural features by examining the radial distribution functions of liquid $Sb_2Te$ and $Sb_2Te_3$ at 1100K. The simulations were performed using 180-atom cubic supercells (with lattice constants a=b=c=18.75Å and α=β=γ=90°). Each system was first melted at 2000 K, followed by equilibration in the NVT ensemble at 1100 K for 20 ps. The RDFs were computed using the last 1000 frames of trajectory data.

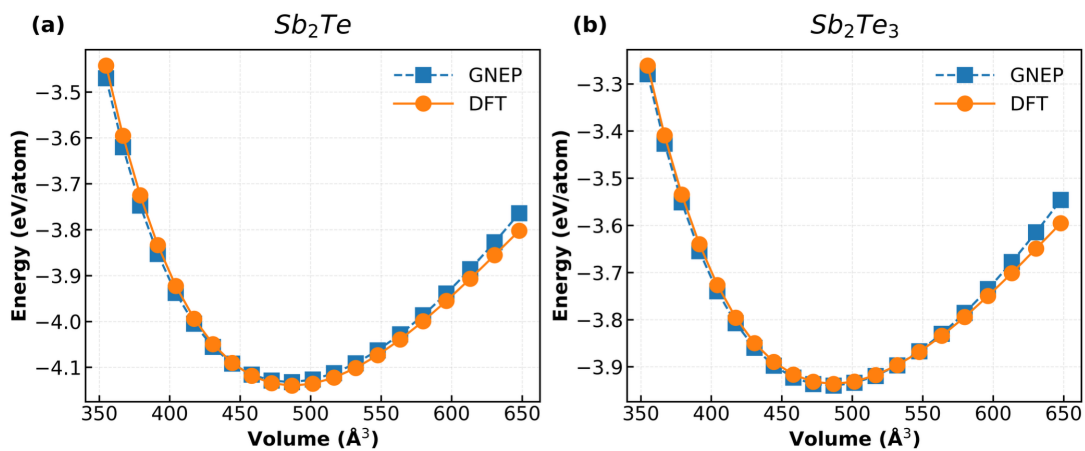

**Figure 5** The EOS validation for crystalline Sb–Te structures. (a) $P\bar{3}m1$ $Sb_2Te$. (b) $R\bar{3}m$ $Sb_2Te_3$.

As shown in **Figure 6**, the molecular dynamics simulations using the GNEP-trained potential are in good agreement with reference AIMD data. The model successfully reproduces the main structural features of the liquid phases. A minor overestimation of structural order is observed near 4 Å in liquid $Sb_2Te_3$. This difference was also observed with our previous study[8], but the structural characteristics of the main peak remained consistent.

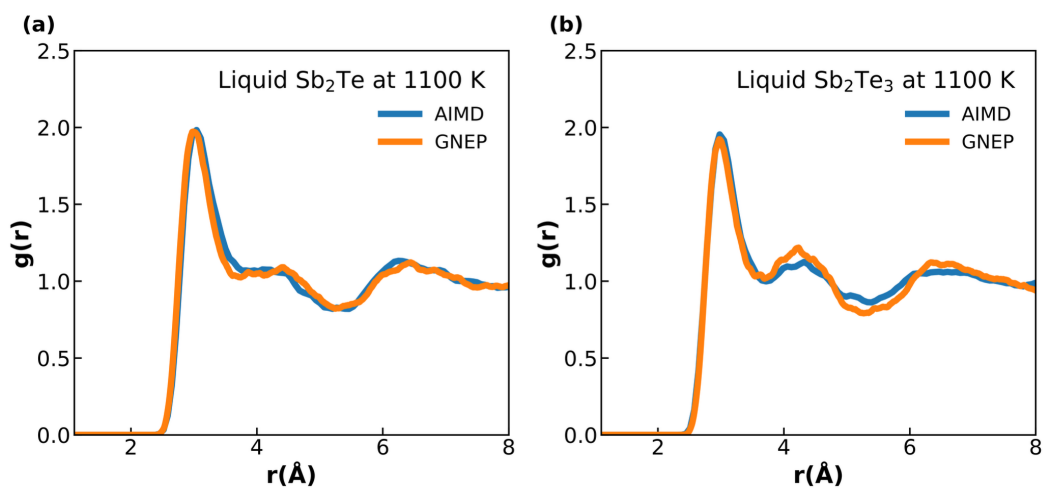

**Figure 6** The radial distribution function (RDF) validation for liquid Sb–Te compounds at 1100 K. (a) liquid $Sb_2Te$. (b) liquid $Sb_2Te_3$.

These evaluations confirm that the GNEP potential not only achieves excellent predictive accuracy but also exhibits strong generalization ability and reliable physical predictions. Its low computational cost makes it well-suited for large-scale molecular dynamics simulations, enabling efficient exploration of complex material behaviors.

## IV. CONCLUSION

In this work, we developed a gradient-based training framework for Neuroevolution Potentials by introducing explicit analytical gradients and leveraging the Adam optimizer. Compared to the original SNES-based approach, GNEP dramatically improves training efficiency—reducing the required number of epochs by an order of magnitude—while maintaining accuracy, generalization, and physical consistency. The resulting interatomic potential accurately reproduces energy–volume curves and radial distribution functions in Sb-Te systems, showing closely agreement with DFT and AIMD reference data. These results demonstrate that the GNEP approach offers a reliable and efficient alternative to evolutionary training strategies, particularly for large-scale, high-accuracy simulations.

Nevertheless, implementing analytic gradients, especially for the custom polynomial-based descriptors employed in NEP, introduces substantial complexity compared to using a black-box optimizer. In our CUDA implementation, special care was needed to handle differentiation of the descriptors, a process that is both technically demanding and prone to error. Any future modifications to the descriptor definitions would necessitate re-deriving and re-implementing the corresponding gradient expressions. By contrast, one significant advantage of SNES was its flexibility: modifications to the model architecture could be made without requiring any gradient-related changes. Consequently, although the adoption of analytic gradients improves training efficiency, it also increases the maintenance burden and reduces flexibility.

**Code Availability**

The GNEP training framework developed in this work is open source and available at https://github.com/hfood02/GNEP. Molecular dynamics simulations were performed using the GPUMD package, available at https://github.com/brucefan1983/GPUMD.

## ACKNOWLEDGMENTS

This work was supported by the National Key Research and Development Program of China (2022YFB3807200) and the National Natural Science Foundation of China (grant No. 52332005).

**APPENDIX A: Chebyshev Polynomials and Cutoff Function**

1. Chebyshev Polynomials of the First Kind

In NEP, the radial descriptors are constructed using Chebyshev polynomials of the first kind $T_k(\xi)$, which form a complete set of orthogonal basis functions on the interval $[-1,1]$. These polynomials are recursively defined as follows:

$$T_0(\xi) = 1, \quad T_1(\xi) = \xi, \quad T_k(\xi) = 2\xi T_{k-1}(\xi) - T_{k-2}(\xi), \quad k \geq 2 \quad (A1.1)$$

The argument $\xi$ is a scaled variable of the interatomic distance $r_{ij}$, mapped from the range $[0, r_c]$ to $[-1,1]$ via:

$$\xi = 2\left(\frac{r_{ij}}{r_c} - 1\right)^2 - 1 \quad (A1.2)$$

The derivatives of the Chebyshev polynomials required for force and virial gradient calculations are computed recursively as:

$$\frac{\partial T_0(\xi)}{\partial \xi} = 0 \quad (A1.3)$$

$$\frac{\partial T_1(\xi)}{\partial \xi} = 1 \quad (A1.4)$$

$$\frac{\partial T_k(\xi)}{\partial \xi} = 2T_{k-1}(\xi) + 2\xi \frac{\partial T_{k-1}(\xi)}{\partial \xi} - \frac{\partial T_{k-2}(\xi)}{\partial \xi}, \quad k \geq 2 \quad (A1.5)$$

2. Smooth Cutoff Function

To smoothly suppress contributions from distant atoms, a cutoff function $f_c(r_{ij})$ is introduced:

$$f_c(r_{ij}) = \begin{cases} \frac{1}{2}\left[1 + \cos\left(\pi \frac{r_{ij}}{r_c}\right)\right], & \text{if } r_{ij} \leq r_c \\ 0, & \text{if } r_{ij} > r_c \end{cases} \quad (A2.1)$$

This function is equal to 1 at $r_{ij} = 0$, and smoothly decays to 0 at $r_{ij} = r_c$, ensuring continuous values and first derivatives. Its derivative is:

$$f_c'(r_{ij}) = \frac{\partial f_c(r_{ij})}{\partial r_{ij}} = \begin{cases} -\frac{\pi}{2r_c} \sin\left(\pi \frac{r_{ij}}{r_c}\right), & \text{if } r_{ij} \leq r_c \\ 0, & \text{if } r_{ij} > r_c \end{cases} \quad (A2.2)$$

The use of the cutoff function ensures the continuity of both the potential energy and its gradients at the cutoff boundary, avoiding unphysical singularities. Together, the Chebyshev polynomial basis and cutoff function constitute the mathematical foundation for constructing smooth and symmetric local atomic descriptors.

**APPENDIX B: Real Spherical Harmonics Basis**

The angular part of the three-body descriptor is constructed using real spherical harmonics, expressed in terms of Cartesian coordinates. These functions form an orthonormal basis over the unit sphere and allow encoding of angular information in a way that is invariant to global rotations.

The real spherical harmonics basis functions $b_{lm}(x_{ij}, y_{ij}, z_{ij})$ used in NEP follow the convention:

$$b_{10} = z_{ij} \tag{B1.1}$$

$$b_{11} = x_{ij} \tag{B1.2}$$

$$b_{12} = y_{ij} \tag{B1.3}$$

$$b_{20} = 3z_{ij}^2 - r_{ij}^2 \tag{B2.1}$$

$$b_{21} = x_{ij} z_{ij} \tag{B2.2}$$

$$b_{22} = y_{ij} z_{ij} \tag{B2.3}$$

$$b_{23} = x_{ij}^2 - y_{ij}^2 \tag{B2.4}$$

$$b_{24} = 2 x_{ij} y_{ij} \tag{B2.5}$$

$$b_{30} = 5z_{ij}^3 - 3z_{ij} r_{ij}^2 \tag{B3.1}$$

$$b_{31} = (5z_{ij}^2 - r_{ij}^2) x_{ij} \tag{B3.2}$$

$$b_{32} = (5z_{ij}^2 - r_{ij}^2) y_{ij} \tag{B3.3}$$

$$b_{33} = (x_{ij}^2 - y_{ij}^2) z_{ij} \tag{B3.4}$$

$$b_{34} = 2 x_{ij} y_{ij} z_{ij} \tag{B3.5}$$

$$b_{35} = (x_{ij}^2 - 3y_{ij}^2) x_{ij} \tag{B3.6}$$

$$b_{36} = (3x_{ij}^2 - y_{ij}^2) y_{ij} \tag{B3.7}$$

$$b_{40} = 35 z_{ij}^4 - 30 z_{ij}^2 r_{ij}^2 + 3 r_{ij}^4 \tag{B4.1}$$

$$b_{41} = (7z_{ij}^3 - 3z_{ij}r_{ij}^2)x_{ij} \quad (B4.2)$$

$$b_{42} = (7z_{ij}^3 - 3z_{ij}r_{ij}^2)y_{ij} \quad (B4.3)$$

$$b_{43} = (7z_{ij}^2 - r_{ij}^2)(x_{ij}^2 - y_{ij}^2) \quad (B4.4)$$

$$b_{44} = (7z_{ij}^2 - r_{ij}^2)2x_{ij}y_{ij} \quad (B4.5)$$

$$b_{45} = (x_{ij}^3 - 3x_{ij}y_{ij}^2)z_{ij} \quad (B4.6)$$

$$b_{46} = (3x_{ij}^2 y_{ij} - y_{ij}^3)z_{ij} \quad (B4.7)$$

$$b_{47} = x_{ij}^4 - 6x_{ij}^2 y_{ij}^2 + y_{ij}^4 \quad (B4.8)$$

$$b_{48} = 4x_{ij}^3 y_{ij} - 4x_{ij}y_{ij}^3 \quad (B4.9)$$

$$b_{50} = 63z_{ij}^5 - 70z_{ij}^3 r_{ij}^2 + 15z_{ij}r_{ij}^4 \quad (B5.1)$$

$$b_{51} = (21z_{ij}^4 - 14z_{ij}^2 r_{ij}^2 + r_{ij}^4)x_{ij} \quad (B5.2)$$

$$b_{52} = (21z_{ij}^4 - 14z_{ij}^2 r_{ij}^2 + r_{ij}^4)y_{ij} \quad (B5.3)$$

$$b_{53} = (3z_{ij}^3 - z_{ij}r_{ij}^2)(x_{ij}^2 - y_{ij}^2) \quad (B5.4)$$

$$b_{54} = (3z_{ij}^3 - z_{ij}r_{ij}^2)2x_{ij}y_{ij} \quad (B5.5)$$

$$b_{55} = (9z_{ij}^2 - r_{ij}^2)(x_{ij}^3 - 3x_{ij}y_{ij}^2) \quad (B5.6)$$

$$b_{56} = (9z_{ij}^2 - r_{ij}^2)(3x_{ij}^2 y_{ij} - y_{ij}^3) \quad (B5.7)$$

$$b_{57} = z_{ij}(x_{ij}^4 - 6x_{ij}^2 y_{ij}^2 + y_{ij}^4) \quad (B5.8)$$

$$b_{58} = z_{ij}(4x_{ij}^3 y_{ij} - 4x_{ij}y_{ij}^3) \quad (B5.9)$$

$$b_{59} = x_{ij}^5 - 10x_{ij}^3 y_{ij}^2 + 5x_{ij}y_{ij}^4 \quad (B5.10)$$

$$b_{510} = 5x_{ij}^4 y_{ij} - 10x_{ij}^2 y_{ij}^3 + y_{ij}^5 \qquad (B5.11)$$

$$b_{60} = 231z_{ij}^6 - 315z_{ij}^4 r_{ij}^2 + 105z_{ij}^2 r_{ij}^4 - 5r_{ij}^6 \qquad (B6.1)$$

$$b_{61} = \left(33z_{ij}^5 - 30z_{ij}^3 r_{ij}^2 + 5z_{ij} r_{ij}^4\right)x_{ij} \qquad (B6.2)$$

$$b_{62} = \left(33z_{ij}^5 - 30z_{ij}^3 r_{ij}^2 + 5z_{ij} r_{ij}^4\right)y_{ij} \qquad (B6.3)$$

$$b_{63} = \left(33z_{ij}^4 - 18z_{ij}^2 r_{ij}^2 + r_{ij}^4\right)\left(x_{ij}^2 - y_{ij}^2\right) \qquad (B6.4)$$

$$b_{64} = \left(33z_{ij}^4 - 18z_{ij}^2 r_{ij}^2 + r_{ij}^4\right)2x_{ij}y_{ij} \qquad (B6.5)$$

$$b_{65} = \left(11z_{ij}^3 - 3z_{ij} r_{ij}^2\right)\left(x_{ij}^3 - 3x_{ij}y_{ij}^2\right) \qquad (B6.6)$$

$$b_{66} = \left(11z_{ij}^3 - 3z_{ij} r_{ij}^2\right)\left(3x_{ij}^2 y_{ij} - y_{ij}^3\right) \qquad (B6.7)$$

$$b_{67} = \left(11z_{ij}^2 - r_{ij}^2\right)\left(x_{ij}^4 - 6x_{ij}^2 y_{ij}^2 + y_{ij}^4\right) \qquad (B6.8)$$

$$b_{68} = \left(11z_{ij}^2 - r_{ij}^2\right)\left(4x_{ij}^3 y_{ij} - 4x_{ij}y_{ij}^3\right) \qquad (B6.9)$$

$$b_{69} = z_{ij}\left(x_{ij}^5 - 10x_{ij}^3 y_{ij}^2 + 5x_{ij}y_{ij}^4\right) \qquad (B6.10)$$

$$b_{610} = z_{ij}\left(5x_{ij}^4 y_{ij} - 10x_{ij}^2 y_{ij}^3 + y_{ij}^5\right) \qquad (B6.11)$$

$$b_{611} = x_{ij}^6 - 15x_{ij}^4 y_{ij}^2 + 15x_{ij}^2 y_{ij}^4 - y_{ij}^6 \qquad (B6.12)$$

$$b_{612} = 6x_{ij}^5 y_{ij} - 20x_{ij}^3 y_{ij}^3 + 6x_{ij}y_{ij}^5 \qquad (B6.13)$$

$$b_{70} = 429z_{ij}^7 - 693z_{ij}^5 r_{ij}^2 + 315z_{ij}^3 r_{ij}^4 - 35z_{ij} r_{ij}^6 \qquad (B7.1)$$

$$b_{71} = \left(429z_{ij}^6 - 495z_{ij}^4 r_{ij}^2 + 135z_{ij}^2 r_{ij}^4 - 5r_{ij}^6\right)x_{ij} \qquad (B7.2)$$

$$b_{72} = \left(429z_{ij}^6 - 495z_{ij}^4 r_{ij}^2 + 135z_{ij}^2 r_{ij}^4 - 5r_{ij}^6\right)y_{ij} \qquad (B7.3)$$

$$b_{73} = \left(143z_{ij}^5 - 110z_{ij}^3 r_{ij}^2 + 15z_{ij} r_{ij}^4\right)\left(x_{ij}^2 - y_{ij}^2\right) \qquad (B7.4)$$

$$b_{74} = (143z_{ij}^5 - 110z_{ij}^3 r_{ij}^2 + 15z_{ij} r_{ij}^4) 2x_{ij} y_{ij} \tag{B7.5}$$

$$b_{75} = (143z_{ij}^4 - 66z_{ij}^2 r_{ij}^2 + 3r_{ij}^4)(x_{ij}^3 - 3x_{ij} y_{ij}^2) \tag{B7.6}$$

$$b_{76} = (143z_{ij}^4 - 66z_{ij}^2 r_{ij}^2 + 3r_{ij}^4)(3x_{ij}^2 y_{ij} - y_{ij}^3) \tag{B7.7}$$

$$b_{77} = (13z_{ij}^3 - 3z_{ij} r_{ij}^2)(x_{ij}^4 - 6x_{ij}^2 y_{ij}^2 + y_{ij}^4) \tag{B7.8}$$

$$b_{78} = (13z_{ij}^3 - 3z_{ij} r_{ij}^2)(4x_{ij}^3 y_{ij} - 4x_{ij} y_{ij}^3) \tag{B7.9}$$

$$b_{79} = (13z_{ij}^2 - r_{ij}^2)(x_{ij}^5 - 10x_{ij}^3 y_{ij}^2 + 5x_{ij} y_{ij}^4) \tag{B7.10}$$

$$b_{710} = (13z_{ij}^2 - r_{ij}^2)(5x_{ij}^4 y_{ij} - 10x_{ij}^2 y_{ij}^3 + y_{ij}^5) \tag{B7.11}$$

$$b_{711} = z_{ij}(x_{ij}^6 - 15x_{ij}^4 y_{ij}^2 + 15x_{ij}^2 y_{ij}^4 - y_{ij}^6) \tag{B7.12}$$

$$b_{712} = z_{ij}(6x_{ij}^5 y_{ij} - 20x_{ij}^3 y_{ij}^3 + 6x_{ij} y_{ij}^5) \tag{B7.13}$$

$$b_{713} = x_{ij}^7 - 21x_{ij}^5 y_{ij}^2 + 35x_{ij}^3 y_{ij}^4 - 7x_{ij} y_{ij}^6 \tag{B7.14}$$

$$b_{714} = 7x_{ij}^6 y_{ij} - 35x_{ij}^4 y_{ij}^3 + 21x_{ij}^2 y_{ij}^5 - y_{ij}^7 \tag{B7.15}$$

$$b_{80} = 6435z_{ij}^8 - 12012z_{ij}^6 r_{ij}^2 + 6930z_{ij}^4 r_{ij}^4 - 1260z_{ij}^2 r_{ij}^6 + 35r_{ij}^8 \tag{B8.1}$$

$$b_{81} = (715z_{ij}^7 - 1001z_{ij}^5 r_{ij}^2 + 385z_{ij}^3 r_{ij}^4 - 35z_{ij} r_{ij}^6) x_{ij} \tag{B8.2}$$

$$b_{82} = (715z_{ij}^7 - 1001z_{ij}^5 r_{ij}^2 + 385z_{ij}^3 r_{ij}^4 - 35z_{ij} r_{ij}^6) y_{ij} \tag{B8.3}$$

$$b_{83} = (143z_{ij}^6 - 143z_{ij}^4 r_{ij}^2 + 33z_{ij}^2 r_{ij}^4 - r_{ij}^6)(x_{ij}^2 - y_{ij}^2) \tag{B8.4}$$

$$b_{84} = (143z_{ij}^6 - 143z_{ij}^4 r_{ij}^2 + 33z_{ij}^2 r_{ij}^4 - r_{ij}^6) 2x_{ij} y_{ij} \tag{B8.5}$$

$$b_{85} = (39z_{ij}^5 - 26z_{ij}^3 r_{ij}^2 + 3z_{ij} r_{ij}^4)(x_{ij}^3 - 3x_{ij} y_{ij}^2) \tag{B8.6}$$

$$b_{86} = (39z_{ij}^5 - 26z_{ij}^3 r_{ij}^2 + 3z_{ij} r_{ij}^4)(3x_{ij}^2 y_{ij} - y_{ij}^3) \tag{B8.7}$$

$$b_{87} = (65z_{ij}^4 - 26z_{ij}^2 r_{ij}^2 + r_{ij}^4)(x_{ij}^4 - 6x_{ij}^2 y_{ij}^2 + y_{ij}^4) \qquad (B8.8)$$

$$b_{88} = (65z_{ij}^4 - 26z_{ij}^2 r_{ij}^2 + r_{ij}^4)(4x_{ij}^3 y_{ij} - 4x_{ij} y_{ij}^3) \qquad (B8.9)$$

$$b_{89} = (5z_{ij}^3 - z_{ij} r_{ij}^2)(x_{ij}^5 - 10x_{ij}^3 y_{ij}^2 + 5x_{ij} y_{ij}^4) \qquad (B8.10)$$

$$b_{810} = (5z_{ij}^3 - z_{ij} r_{ij}^2)(5x_{ij}^4 y_{ij} - 10x_{ij}^2 y_{ij}^3 + y_{ij}^5) \qquad (B8.11)$$

$$b_{811} = (15z_{ij}^2 - r_{ij}^2)(x_{ij}^6 - 15x_{ij}^4 y_{ij}^2 + 15x_{ij}^2 y_{ij}^4 - y_{ij}^6) \qquad (B8.12)$$

$$b_{812} = (15z_{ij}^2 - r_{ij}^2)(6x_{ij}^5 y_{ij} - 20x_{ij}^3 y_{ij}^3 + 6x_{ij} y_{ij}^5) \qquad (B8.13)$$

$$b_{813} = z_{ij}(x_{ij}^7 - 21x_{ij}^5 y_{ij}^2 + 35x_{ij}^3 y_{ij}^4 - 7x_{ij} y_{ij}^6) \qquad (B8.14)$$

$$b_{814} = z_{ij}(7x_{ij}^6 y_{ij} - 35x_{ij}^4 y_{ij}^3 + 21x_{ij}^2 y_{ij}^5 - y_{ij}^7) \qquad (B8.15)$$

$$b_{815} = x_{ij}^8 - 28x_{ij}^6 y_{ij}^2 + 70x_{ij}^4 y_{ij}^4 - 28x_{ij}^2 y_{ij}^6 + y_{ij}^8 \qquad (B8.16)$$

$$b_{816} = 8x_{ij}^7 y_{ij} - 56x_{ij}^5 y_{ij}^3 + 56x_{ij}^3 y_{ij}^5 - 8x_{ij} y_{ij}^7 \qquad (B8.17)$$

Each basis function is further scaled by $r_{ij}^l$ to normalize the angular dependence in the construction of descriptors:

$$\boldsymbol{B}_{lm} = \frac{b_{lm}(x_{ij}, y_{ij}, z_{ij})}{r_{ij}^l} \qquad (B9.1)$$

In force and virial calculations, the gradient of $\boldsymbol{B}_{lm}$ with respect to the relative position vector $\boldsymbol{r}_{ij}$ is required:

$$\boldsymbol{B}_{lm}^{\prime(r_{ij})} = \frac{\partial \big(b_{lm}(x_{ij}, y_{ij}, z_{ij})/r_{ij}^l\big)}{\partial \boldsymbol{r}_{ij}} = \left(\frac{\partial \boldsymbol{B}_{lm}}{\partial x_{ij}}, \frac{\partial \boldsymbol{B}_{lm}}{\partial y_{ij}}, \frac{\partial \boldsymbol{B}_{lm}}{\partial z_{ij}}\right) \qquad (B9.2)$$

These derivatives are evaluated analytically and used in the descriptor gradient expressions in Appendix C.

## APPENDIX C: Gradients

1. Radial Descriptor Gradients

For a system consisting of $N$ atoms, where each atom can have up to $M$ neighbors, we now derive the gradients of radial descriptors with respect to atomic positions and expansion coefficients. The gradient of the radial descriptor with respect to the relative position vector $r_{ij}$:

$$Q_n'^{(r_{ij})} = \left\{\frac{\partial q_n^i}{\partial r_{ij}}\right\} \in \mathbb{R}^{N \times M \times 3} = \frac{1}{2} \sum_{k=0}^{N_{bas}^R} c_{nk}^{IJ} f_k'(r_{ij}) \quad (C1.1)$$

The virial-related gradient term involves an additional outer product with $r_{ij}$:

$$Q_n'^{(r_{ji} \otimes r_{ij})} = \left\{\frac{\partial q_n^i}{\partial r_{ji}} r_{ij}\right\} \in \mathbb{R}^{N \times M \times 6} = \frac{1}{2} \sum_{k=0}^{N_{bas}^R} c_{nk}^{IJ} f_k'(r_{ji}) r_{ij} \quad (C1.2)$$

During training, the expansion coefficients $c_{nk}^{IJ}$ are trainable parameters. To support gradient-based optimization, we define their first and second derivatives.

(a) First-order gradient with respect to parameters:

$$Q_n'^{(c)} = \left\{\frac{\partial q_n^i}{\partial c_{nk}^{IJ}}\right\} \in \mathbb{R}^{N \times M \times N_{bas}^R} = \frac{T_k(\xi) + 1}{2} f_c(r_{ij}) \quad (C1.3)$$

(b) Mixed second-order gradient with respect to parameters and positions:

$$Q_n''^{(c)} = \left\{\frac{\partial^2 q_n^i}{\partial c_{nk}^{IJ} \partial r_{ij}}\right\} \in \mathbb{R}^{N \times M \times 3 \times N_{bas}^R} = f_k'(r_{ij}) \quad (C1.4)$$

$$Q_n''^{(c \otimes r_{ij})} = \left\{\frac{\partial^2 q_n^i}{\partial c_{nk}^{IJ} \partial r_{ji}} r_{ij}\right\} \in \mathbb{R}^{N \times M \times 6 \times N_{bas}^R} = f_k'(r_{ij}) r_{ij} \quad (C1.5)$$

Note that the difference between radial functions ($n$ is the same, k is different) is only reflected in the expansion coefficient.

2. Angular Descriptor Gradients

This section provides full derivations of gradients for angular descriptors. Recall the Equations (7,8) and Equations (B1-9), we can derive:

$$Q_{nl}'^{(r_{ij})} = \left\{\frac{\partial q_{nl}^i}{\partial r_{ij}}\right\} \in \mathbb{R}^{N \times M \times 3} = 2 \sum_{m=0}^{2l} C_{lm} \cdot S_{nlm} S_{nlm}'^{(r_{ij})} \quad (C2.1)$$

The corresponding virial-related term is:

$$Q_{nl}^{\prime(r_{ji}\otimes r_{ij})} = \left\{\frac{\partial q_{nl}^i}{\partial r_{ji}} r_{ij}\right\} \in \mathbb{R}^{N\times M\times 6} = 2\sum_{m=0}^{2l} C_{lm}\cdot S_{nlm}S_{nlm}^{\prime(r_{ji}\otimes r_{ij})} \qquad (C2.2)$$

Where:

$$S_{nlm}^{\prime(r_{ij})} = \left\{\frac{\partial S_{nlm}^i}{\partial r_{ij}}\right\} \in \mathbb{R}^{N\times M\times 3} = g_n(r_{ij})\cdot B_{lm}^{\prime(r_{ij})} + \frac{\partial g_n(r_{ij})}{\partial r_{ij}}\cdot B_{lm} \qquad (C2.3)$$

$$S_{nlm}^{\prime(r_{ji}\otimes r_{ij})} = \left\{\frac{\partial S_{nlm}^i}{\partial r_{ji}} r_{ij}\right\} \in \mathbb{R}^{N\times M\times 6} = S_{nlm}^{\prime(r_{ji})} r_{ij} \qquad (C2.4)$$

Similarly, the gradient of angular expansion coefficients can be derived:

$$Q_{nl}^{\prime(c)} = \left\{\frac{\partial q_{nl}^i}{\partial c_{nk}^{IJ}}\right\} \in \mathbb{R}^{N\times M\times N_{bas}^A} = 2\sum_{m=0}^{2l} C_{lm}\cdot S_{nlm}S_{nlm}^{\prime(c)} \qquad (C2.5)$$

$$Q_{nl}^{\prime\prime(c)} = \left\{\frac{\partial^2 q_{nl}^i}{\partial c_{nk}^{IJ}\partial r_{ij}}\right\} \in \mathbb{R}^{N\times M\times N_{bas}^A\times 3} = 2\sum_{m=0}^{2l} C_{lm}\cdot \left(S_{nlm}S_{nlm}^{\prime\prime(c)} + S_{nlm}^{\prime(c)}\sum_{j\neq i} S_{nlm}^{\prime(r_{ij})}\right) \qquad (C2.6)$$

$$Q_{nl}^{\prime\prime(c\otimes r_{ij})} = \left\{\frac{\partial^2 q_{nl}^i}{\partial c_{nk}^{IJ}\partial r_{ji}} r_{ij}\right\} \in \mathbb{R}^{N\times M\times N_{bas}^A\times 6}$$

$$= 2\sum_{m=0}^{2l} C_{lm}\cdot \left(S_{nlm}S_{nlm}^{\prime\prime(c\otimes r_{ij})} + S_{nlm}^{\prime(c)}\sum_{j\neq i} S_{nlm}^{\prime(r_{ji}\otimes r_{ij})}\right) \qquad (C2.7)$$

Where:

$$S_{nlm}^{\prime(c)} = \left\{\frac{\partial S_{nlm}^i}{\partial c_{nk}^{IJ}}\right\} \in \mathbb{R}^{N\times M\times N_{bas}^A} = \left(\frac{\partial g_n(r_{ij})}{\partial c_{nk}^{IJ}}\frac{b_{lm}(x_{ij},y_{ij},z_{ij})}{r_{ij}^l}\right) = f_k(r_{ij})\cdot B_{lm} \qquad (C2.8)$$

$$S_{nlm}^{\prime\prime(c)} = \left\{\frac{\partial^2 S_{nlm}^i}{\partial c_{nk}^{IJ}\partial r_{ij}}\right\} \in \mathbb{R}^{N\times M\times N_{bas}^A\times 3}$$

$$= \left(\frac{\partial g_n(r_{ij})}{\partial c_{nk}^{IJ}}\frac{\partial b_{lm}(x_{ij},y_{ij},z_{ij})/r_{ij}^l}{\partial r_{ij}}\right) + \left(\frac{\partial^2 g_n(r_{ij})}{\partial c_{nk}^{IJ}\partial r_{ij}}\frac{b_{lm}(x_{ij},y_{ij},z_{ij})}{r_{ij}^l}\right)$$

$$= f_k(r_{ij})\cdot B_{lm}^{\prime(r_{ij})} + f_k'(r_{ij})\cdot B_{lm} \qquad (C2.9)$$

$$S_{nlm}^{\prime\prime(c\otimes r_{ij})} = \left\{\frac{\partial^2 S_{nlm}^i}{\partial c_{nk}^{IJ}\partial r_{ji}} r_{ij}\right\} \in \mathbb{R}^{N\times M\times N_{bas}^A\times 6} = S_{nlm}^{\prime\prime(c)} r_{ij} \qquad (C2.10)$$

It is important to note that the expansion coefficients $c_{nk}^{IJ}$ explicitly incorporate the element types of the central atom $i$ and its neighboring atom $j$. However, the three-body angular descriptor components $S_{nlm}$ and their derivatives involve summation over neighboring atoms. Therefore, the specific mapping of element types must be determined in conjunction with the enumeration process of neighbor pairs, as detailed in the computation flow shown in $Algorithm$ 1.

---

**Algorithm 1**: calculate $Q_{nl}^{''(c)}[type[i], type[j]]$

---

**Input**: $\{i\}, \{j\}, \{type\}$
1. **for** $i = 0, 1, \ldots, N$ **do**
2.    $t_1 = type[i]$
3.    **for** $j = 0, 1, \ldots, M$ **do**
4.      $t_2 = type[j]$
5.      $r_{ij} = position[j] - position[i]$
6.      $Q_{nl}^{''(c)}(t_1, t_2) += S_{nlm}(r_{ij}, t_1) \cdot S_{nlm}^{''(c)}(r_{ij}, t_1, t_2)$
7.      $S_{nlm}^{'(r_{ij})} = S_{nlm}^{'(r_{ij})}(r_{ij}, t_1)$
8.      **for** $j' = 0, 1, \ldots, M$ **do**
9.         $t_2' = type[j']$
10.        $r_{ij}' = position[j'] - position[i]$
11.        $Q_{nl}^{''(c)}(t_1, t_2') += S_{nlm}^{'(c)}(r_{ij}', t_1, t_2') \cdot S_{nlm}^{'(r_{ij})}$
12.      **end for**
13.    **end for**
14. **end for**

**Output**: $Q_{nl}^{''(c)}$

---

3. Gradients of Energy, Force, and Virial Stress

This section summarizes how energy, atomic forces, and virial stress are computed analytically based on the gradients of local atomic descriptors with respect to atomic positions and model parameters.

(a) Energy gradients

The atomic descriptor vector $\boldsymbol{G_i} = \{q_n^i, q_{nl}^i\} \in \mathbb{R}^{N_{des}}, N_{des} = N_q^R + N_q^A$ includes both radial and angular descriptors. Each atomic energy is given by $E_i = \mathcal{F}(\boldsymbol{G_i}) \in \mathbb{R}^1$. The gradient of energy with respect to descriptors:

$$E'_i = \frac{\partial \mathcal{F}(G_i)}{\partial G_i} = \begin{bmatrix} \dfrac{\partial E_i}{\partial G_i(q_n^i)} \\ \dfrac{\partial E_i}{\partial G_i(q_{nl}^i)} \end{bmatrix} \in \mathbb{R}^{N_{des}} \tag{C3.1}$$

To consider cross-descriptor coupling, the second-order derivative is:

$$E''_i = \frac{\partial^2 \mathcal{F}(G_i)}{\partial G_{i'} \, \partial G_i} \in \mathbb{R}^{N_{des} \times N_{des}} \tag{C3.2}$$

See **Appendix D** for details of the above first-order derivatives and second-order derivatives.

Splitting descriptor vector by origin (radial/angular) for clarity:

$$G_i : \left( \underbrace{(c_1^R, \ldots, c_n^R)}_{radial}, \underbrace{(c_1^A, \ldots, c_{nl}^A)}_{angular} \right) \tag{C3.3}$$

This reflects how the coupling between different descriptor components affects the energy. Accordingly, the second-order derivative of the energy in Equation ($C3.2$) can be expressed in a block matrix form as follows:

$$E''_i = \begin{bmatrix} \dfrac{\partial^2 \mathcal{F}(G_i)}{\partial G_i(q_n^i) \partial G_i(q_n^i)} & \dfrac{\partial^2 \mathcal{F}(G_i)}{G_i(q_n^i) \partial G_i(q_{nl}^i)} \\ \dfrac{\partial^2 \mathcal{F}(G_i)}{G_i(q_{nl}^i) \partial G_i(q_n^i)} & \dfrac{\partial^2 \mathcal{F}(G_i)}{G_i(q_{nl}^i) \partial G_i(q_{nl}^i)} \end{bmatrix} \tag{C3.4}$$

Based on the chain rule and the previously derived first-order derivative tensors of the descriptors with respect to the expansion coefficients, $Q_n^{\prime(c)}$ and $Q_{nl}^{\prime(c)}$, the energy gradient with respect to the expansion coefficients can be obtained:

$$E_i^{\prime(c)} = \frac{\partial E_i}{\partial c_{nk}^{IJ}} = \frac{\partial E_i}{\partial G_i(q_n^i, q_{nl}^i)} \frac{\partial G_i(q_n^i, q_{nl}^i)}{\partial c_{nk}^{IJ}} = \begin{bmatrix} \dfrac{\partial E_i}{\partial G_i(q_n^i)} \cdot Q_n^{\prime(c)} \\ \dfrac{\partial E_i}{\partial G_i(q_{nl}^i)} \cdot Q_{nl}^{\prime(c)} \end{bmatrix} \in \mathbb{R}^{M \times N_{des} \times N_{bas}} \tag{C3.5}$$

(b) Force gradients

In the weak form of the force given earlier Equation (3), the total force acting on atom $i$:

$$F_i = \sum_{j \in \mathcal{N}^i} \left( \frac{\partial E_i}{\partial r_{ij}} - \frac{\partial E_j}{\partial r_{ji}} \right)$$

With chain rule, we write:

$$\frac{\partial E_i}{\partial r_{ij}} = \sum_{n=0}^{N_q} \frac{\partial E_i}{\partial G_i(q_n^i, q_{nl}^i)} \frac{\partial G_i(q_n^i, q_{nl}^i)}{\partial r_{ij}} = \left\{ \begin{bmatrix} Q_n'^{(r_{ij})} & Q_{nl}'^{(r_{ij})} \end{bmatrix} \times \begin{bmatrix} E_i'(q_n^i) \\ E_i'(q_{nl}^i) \end{bmatrix} \right\} \in \mathbb{R}^{M \times 3} \quad (C3.6)$$

The derivative of partial force with respect to expansion coefficients is given by:

$$\frac{\partial}{\partial c_{nk}^{IJ}} \left( \frac{\partial E_i}{\partial r_{ij}} \right) = \frac{\partial}{\partial c_{nk}^{IJ}} \left( \frac{\partial E_i}{\partial G_i(q_n^i, q_{nl}^i)} \frac{\partial G_i(q_n^i, q_{nl}^i)}{\partial r_{ij}} \right)$$

$$= \frac{\partial E_i}{\partial G_i} \frac{\partial}{\partial c_{nk}^{IJ}} \left( \frac{\partial G_i}{\partial r_{ij}} \right) + \frac{\partial G_i}{\partial r_{ij}} \frac{\partial}{\partial c_{nk}^{IJ}} \left( \frac{\partial E_i}{\partial G_i} \right) \quad (C3.7)$$

In the second term of the above expression, $\frac{\partial E_i}{\partial G_i}$ depends on $G_i$, which in turn is a function of $c_{nk}^{IJ}$. Consequently, the derivative $\frac{\partial}{\partial c_{nk}^{IJ}} \left( \frac{\partial E_i}{\partial G_i} \right)$ can be further expanded using the chain rule:

$$\frac{\partial}{\partial c_{nk}^{IJ}} \left( \frac{\partial E_i}{\partial G_i} \right) = \frac{\partial G_i}{\partial c_{nk}^{IJ}} \frac{\partial}{\partial G_i} \left( \frac{\partial E_i}{\partial G_i} \right) \quad (C3.8)$$

Given that the neural network inputs consist of both radial and angular descriptors, the second-order derivatives of the energy with respect to atomic coordinates naturally decompose into radial and angular components.

Consequently, the gradient of the force with respect to the expansion coefficients must be computed separately for the radial and angular components.

By substituting Equations ($C3.7$) and ($C3.8$) into Equation (3), we derive the expression for the derivative of the radial component of the atom $i$ with respect to the radial expansion coefficients $c^R$:

$$\nabla_{ij}^{R:R} = \frac{\partial}{\partial c_{nk}^{IJ}} \left( \frac{\partial E_i}{\partial r_{ij}} \right) \in \mathbb{R}^{M \times N_q^R \times N_{bas}^R \times 3} = E_i'(n) Q_n''^{(c)}(i,j) + Q_n'^{(c)}(i,j') E_i''(n,n') Q_{n'}'^{(r_{ij})}(i,j)$$

$$= \begin{bmatrix} E_i'(0) \cdot Q_0''^{(c)}(i,0) & E_i'(0) \cdot Q_0''^{(c)}(i,1) & \cdots & E_i'(0) \cdot Q_0''^{(c)}(i,M-1) \\ E_i'(1) \cdot Q_1''^{(c)}(i,0) & E_i'(1) \cdot Q_1''^{(c)}(i,1) & & E_i'(1) \cdot Q_1''^{(c)}(i,M-1) \\ \vdots & & \ddots & \vdots \\ E_i'(N_q^R - 1) \cdot Q_{N_q^R-1}''^{(c)}(i,0) & E_i'(N_q^R - 1) \cdot Q_{N_q^R-1}''^{(c)}(i,1) & \cdots & E_i'(N_q^R - 1) \cdot Q_{N_q^R-1}''^{(c)}(i,M-1) \end{bmatrix}$$

$$+\begin{bmatrix} Q_0'^{(c)}(i,0) & Q_0'^{(c)}(i,1) & \cdots & Q_0'^{(c)}(i,M-1) \\ Q_1'^{(c)}(i,0) & Q_1'^{(c)}(i,1) & & Q_1'^{(c)}(i,M-1) \\ \vdots & & \ddots & \vdots \\ Q_{N_q^R-1}'^{(c)}(i,0) & Q_{N_q^R-1}'^{(c)}(i,1) & \cdots & Q_{N_q^R-1}'^{(c)}(i,M-1) \end{bmatrix} \circ \begin{bmatrix} \sum_{n'=0}^{N_q^R-1}\left(E_i''(0,n')\sum_j Q_{n'}'^{(r_{ij})}(i,j)\right) \\ \sum_{n'=0}^{N_q^R-1}\left(E_i''(1,n')\sum_j Q_{n'}'^{(r_{ij})}(i,j)\right) \\ \vdots \\ \sum_{n'=0}^{N_q^R-1}\left(E_i''(N_q^R-1,n')\sum_j Q_{n'}'^{(r_{ij})}(i,j)\right) \end{bmatrix} \quad (C3.9)$$

Here, ∘ denotes the Hadamard (element-wise) product, and $j' \in \{0,1,2,\ldots,M-1\}$, where $M$ is the number of radial neighbors of the atom $i$.

Similarly, the derivative of the radial component of the neighbor atom $j$ with respect to the radial expansion coefficients $c^R$ is given by:

$$\nabla_{ji}^{R:R} = \frac{\partial}{\partial c_{nk}^{IJ}}\left(\frac{\partial E_j}{\partial r_{ji}}\right) \in \mathbb{R}^{M \times N_q^R \times N_{bas}^R \times 3} \quad (C3.10)$$

Combining both contributions, the total gradient becomes:

$$\left.\frac{\partial F_i}{\partial c_{nk}^{IJ}}\right|_{R:R} \in \mathbb{R}^{M \times N_q^R \times N_{bas}^R \times 3} = \nabla_{ij}^{R:R} - \nabla_{ji}^{R:R} \quad (C3.11)$$

Furthermore, the derivative of the radial component of the atom $i$ with respect to the angular expansion coefficients $c^A$ is given by:

$$\nabla_{ij}^{R:A} = \frac{\partial}{\partial c_{nk}^{IJ}}\left(\frac{\partial E_i}{\partial r_{ij}}\right) \in \mathbb{R}^{M \times N_q^A \times N_{bas}^A \times 3} = Q_{nl}'^{(c)}(i,j')E_i''(nl,n')Q_{n'}'^{(r_{ij})}(i,j)$$

$$=\begin{bmatrix} Q_{N_q^R}'^{(c)}(i,0) & Q_{N_q^R}'^{(c)}(i,1) & \cdots & Q_{N_q^R}'^{(c)}(i,M-1) \\ Q_{N_q^R+1}'^{(c)}(i,0) & Q_{N_q^R+1}'^{(c)}(i,1) & & Q_{N_q^R+1}'^{(c)}(i,M-1) \\ \vdots & & \ddots & \vdots \\ Q_{N_q-1}'^{(c)}(i,0) & Q_{N_q-1}'^{(c)}(i,1) & \cdots & Q_{N_q-1}'^{(c)}(i,M-1) \end{bmatrix} \circ \begin{bmatrix} \sum_{n'=0}^{N_q^R-1}\left(E_i''(N_q^R,n')\sum_j Q_{n'}'^{(r_{ij})}(i,j)\right) \\ \sum_{n'=0}^{N_q^R-1}\left(E_i''(N_q^R+1,n')\sum_j Q_{n'}'^{(r_{ij})}(i,j)\right) \\ \vdots \\ \sum_{n'=0}^{N_q^R-1}\left(E_i''(N_q-1,n')\sum_j Q_{n'}'^{(r_{ij})}(i,j)\right) \end{bmatrix} \quad (C3.12)$$

Here, $j' \in \{0,1,2,\ldots,M-1\}$, where $M$ denotes the number of angular neighbors of the atom $i$.

Accordingly, the gradient of the radial force with respect to the angular expansion coefficients $c^A$ is given by:

$$\left.\frac{\partial F_i}{\partial c_{nk}^{IJ}}\right|_{R:A} \in \mathbb{R}^{M \times N_q^A \times N_{bas}^A \times 3} = \nabla_{ij}^{R:A} - \nabla_{ji}^{R:A} \qquad (C3.13)$$

Similarly, it can be shown that gradients also exist for the angular force component with respect to both radial and angular expansion coefficients.

Specifically, the derivative of the angular component of the atom $i$ with respect to the angular expansion coefficients $c^A$ is given by:

$$\Delta_{ij}^{A:A} = \frac{\partial}{\partial c_{nk}^{IJ}}\left(\frac{\partial E_i}{\partial r_{ij}}\right) \in \mathbb{R}^{M \times N_q^A \times N_{bas}^A \times 3}$$

$$= E_i'(nl)Q_{nl}''^{(c)}(i,j) + Q_{nl}'^{(c)}(i,j')E_i''(nl,nl')Q_{nl'}'^{(r_{ij})}(i,j)$$

$$= \begin{bmatrix} E_i'(N_q^R) \cdot Q_{N_q^R}''^{(c)}(i,0) & E_i'(N_q^R) \cdot Q_{N_q^R}''^{(c)}(i,1) & \cdots & E_i'(N_q^R) \cdot Q_{N_q^R}''^{(c)}(i,M-1) \\ E_i'(N_q^R+1) \cdot Q_{N_q^R+1}''^{(c)}(i,0) & E_i'(N_q^R+1) \cdot Q_{N_q^R+1}''^{(c)}(i,1) & & E_i'(N_q^R+1) \cdot Q_{N_q^R+1}''^{(c)}(i,M-1) \\ \vdots & & \ddots & \vdots \\ E_i'(N_q-1) \cdot Q_{N_q-1}''^{(c)}(i,0) & E_i'(N_q-1) \cdot Q_{N_q-1}''^{(c)}(i,1) & \cdots & E_i'(N_q-1) \cdot Q_{N_q-1}''^{(c)}(i,M-1) \end{bmatrix}$$

$$+ \begin{bmatrix} Q_{N_q^R}'^{(c)}(i,0) & Q_{N_q^R}'^{(c)}(i,1) & \cdots & Q_{N_q^R}'^{(c)}(i,M-1) \\ Q_{N_q^R+1}'^{(c)}(i,0) & Q_{N_q^R+1}'^{(c)}(i,1) & & Q_{N_q^R+1}'^{(c)}(i,M-1) \\ \vdots & & \ddots & \vdots \\ Q_{N_q-1}'^{(c)}(i,0) & Q_{N_q-1}'^{(c)}(i,1) & \cdots & Q_{N_q-1}'^{(c)}(i,M-1) \end{bmatrix} \circ \begin{bmatrix} \sum_{nl'=N_q^R}^{N_q-1}\left(E_i''(N_q^R,nl')\sum_j Q_{nl'}'^{(r_{ij})}(i,j)\right) \\ \sum_{nl'=N_q^R}^{N_q-1}\left(E_i''(N_q^R+1,nl')\sum_j Q_{nl'}'^{(r_{ij})}(i,j)\right) \\ \vdots \\ \sum_{nl'=N_q^R}^{N_q-1}\left(E_i''(N_q-1,nl')\sum_j Q_{nl'}'^{(r_{ij})}(i,j)\right) \end{bmatrix} \quad (C3.14)$$

Here, $j' \in \{0,1,2,\ldots,M-1\}$, where $M$ denotes the number of angular neighbors of the atom $i$.

Accordingly, the gradient of the angular force with respect to the angular expansion coefficients $c^A$ is given by:

$$\left.\frac{\partial F_i}{\partial c_{nk}^{IJ}}\right|_{A:A} \in \mathbb{R}^{M \times N_q^A \times N_{bas}^A \times 3} = \nabla_{ij}^{A:A} - \nabla_{ji}^{A:A} \qquad (C3.15)$$

Similarly, the derivative of the angular component of the atom $i$ with respect to the

radial expansion coefficients $c^R$ is:

$$\Delta_{ij}^{A:R} = \frac{\partial}{\partial c_{nk}^{IJ}}\left(\frac{\partial E_i}{\partial r_{ij}}\right) \in \mathbb{R}^{M \times N_q^R \times N_{bas}^R \times 3} = \boldsymbol{Q}_n^{\prime(c)}(i,j')\boldsymbol{E}_i^{\prime\prime}(n,nl')\boldsymbol{Q}_{nl'}^{\prime(r_{ij})}(i,j)$$

$$\begin{bmatrix} Q_0^{\prime(c)}(i,0) & Q_0^{\prime(c)}(i,1) & \cdots & Q_0^{\prime(c)}(i,M-1) \\ Q_1^{\prime(c)}(i,0) & Q_1^{\prime(c)}(i,1) & \cdots & Q_1^{\prime(c)}(i,M-1) \\ \vdots & & \ddots & \vdots \\ Q_{N_q^R-1}^{\prime(c)}(i,0) & Q_{N_q^R-1}^{\prime(c)}(i,1) & \cdots & Q_{N_q^R-1}^{\prime(c)}(i,M-1) \end{bmatrix} \circ \begin{bmatrix} \sum_{nl'=N_q^R}^{N_q-1}\left(E_i^{\prime\prime}(0,nl')\sum_j Q_{nl'}^{\prime(r_{ij})}(i,j)\right) \\ \sum_{nl'=N_q^R}^{N_q-1}\left(E_i^{\prime\prime}(1,nl')\sum_j Q_{nl'}^{\prime(r_{ij})}(i,j)\right) \\ \vdots \\ \sum_{nl'=N_q^R}^{N_q-1}\left(E_i^{\prime\prime}(N_q^R-1,nl')\sum_j Q_{nl'}^{\prime(r_{ij})}(i,j)\right) \end{bmatrix} \quad (C3.16)$$

Again, $j' \in \{0,1,2,\ldots,M-1\}$, where $M$ is the number of radial neighbors of the atom $i$.

Thus, the corresponding gradient is:

$$\left.\frac{\partial F_i}{\partial c_{nk}^{IJ}}\right|_{A:R} \in \mathbb{R}^{M \times N_q^R \times N_{bas}^R \times 3} = \nabla_{ij}^{A:R} - \nabla_{ji}^{A:R} \quad (C3.17)$$

(c) Virial gradients

Since the virial tensor can be regarded as the tensor product of force and the relative position vector, its derivative with respect to the expansion coefficients can be directly extended from the force derivatives:

$$\frac{\partial V_i}{\partial c_{nk}^{IJ}} = \frac{\partial}{\partial c_{nk}^{IJ}}\left(\frac{\partial E_i}{\partial r_{ji}} \otimes r_{ij}\right) = \frac{\partial E_i}{\partial G_i}\frac{\partial}{\partial c_{nk}^{IJ}}\left(\frac{\partial G_i}{\partial r_{ji}} \otimes r_{ij}\right) + \frac{\partial G_i}{\partial c_{nk}^{IJ}}\frac{\partial}{\partial G_i}\left(\frac{\partial E_i}{\partial G_i}\right)\left(\frac{\partial G_i}{\partial r_{ji}} \otimes r_{ij}\right) \quad (C3.18)$$

The derivative of the virial stress (either radial or angular) with respect to the expansion coefficients can be expressed similarly. For example, the derivative of the angular virial stress with respect to angular expansion coefficients is given by:

$$\left.\frac{\partial V_i}{\partial c_{nk}^{IJ}}\right|_{A:A} \in \mathbb{R}^{M \times N_q^A \times N_{bas}^A \times 6}$$

$$= \boldsymbol{E}_i'(nl)\boldsymbol{Q}_{nl}^{\prime\prime(c \otimes r_{ij})}(i,j) + \boldsymbol{Q}_{nl}^{\prime(c)}(i,j')\boldsymbol{E}_i^{\prime\prime}(nl,nl')\boldsymbol{Q}_{nl'}^{\prime(r_{ji} \otimes r_{ij})}(i,j) \quad (C3.19)$$

Here, $j' \in \{0,1,2,\ldots,M-1\}$, where $M$ is the number of angular neighbors of the atom $i$.

Likewise, the derivative of the angular virial stress with respect to radial expansion coefficients is given by:

$$\left.\frac{\partial V_i}{\partial c_{nk}^{IJ}}\right|_{A:R} \in \mathbb{R}^{M \times N_q^R \times N_{bas}^R \times 6} = Q_n'^{(c)}(i,j') E_i''(n,nl') Q_{nl'}'^{(r_{ji} \otimes r_{ij})}(i,j) \qquad (C3.20)$$

Where $j' \in \{0,1,2,\dots,M-1\}$, where $M$ is the number of radial neighbors of the atom $i$.

## APPENDIX D: Forward and Backward in the Neural Network

This section derives the forward and backward computations of a single-hidden-layer neural network used to map atomic descriptors $G_i$ to energy.

1. Forward and Backward

For a neural network with $N_{neu}$ hidden neurons, the output energy is:

$$E_i = \mathcal{F}(G_i) = \sum_{\mu=1}^{N_{neu}} w_\mu^{(1)} \tanh\left(\sum_{v=1}^{N_{des}} w_{\mu v}^{(0)} G_{i,v} + b_\mu^{(0)}\right) + b^{(1)} \qquad (D1.1)$$

Define:

$$z_\mu = \sum_{v=1}^{N_{des}} w_{\mu v}^{(0)} G_{i,v} + b_\mu^{(0)} \qquad (D1.2)$$

$$x_\mu = \tanh\left(\sum_{v=1}^{N_{des}} w_{\mu v}^{(0)} G_{i,v} + b_\mu^{(0)}\right) \qquad (D1.3)$$

Then the derivative of activation with respect to descriptor component is:

$$\frac{\partial x_\mu}{\partial G_{i,v}} = (1 - x_\mu^2) w_{\mu v}^{(0)} \qquad (D1.4)$$

The weight connecting input node $v$ to hidden node $\mu$ is denoted as $w_{\mu v}^{(0)}$, with $b_\mu^{(0)}$ representing the bias of the hidden neuron. The weights between the hidden layer and the output node are denoted by $w_\mu^{(1)}$, and $b^{(1)}$ is the bias of the output layer.

The expressions below represent the gradients of the neural network with respect to its inputs ($\frac{\partial \mathcal{F}(G_i)}{\partial G_i}$, $\frac{\partial^2 \mathcal{F}(G_i)}{\partial G_{i\prime} \partial G_i}$) and parameters $p\left(w_{\mu v}^{(0)}, w_{\mu}^{(1)}, b_{\mu}^{(0)}, b^{(1)}\right)$, as obtained via backpropagation:

$$\frac{\partial \mathcal{F}(G_i)}{\partial G_i} = \sum_{\mu=1}^{N_{neu}} w_{\mu}^{(1)} \frac{\partial x_{\mu}}{\partial G_{i,v}} \quad (D1.5)$$

$$\frac{\partial^2 \mathcal{F}(G_i)}{\partial G_{i\prime} \partial G_i} = \frac{\partial}{\partial G_{i,v\prime}} \left(\frac{\partial \mathcal{F}(G_i)}{\partial G_{i,v}}\right) = \sum_{\mu=1}^{N_{neu}} w_{\mu}^{(1)} w_{\mu v}^{(0)} \frac{\partial}{\partial G_{i,v\prime}}(1 - x_{\mu}^2)$$

$$= \sum_{\mu=1}^{N_{neu}} w_{\mu}^{(1)} w_{\mu v}^{(0)} w_{\mu v\prime}^{(0)} \left(-2x_{\mu}(1 - x_{\mu}^2)\right) \quad (D1.6)$$

$$\frac{\partial \mathcal{F}(G_i)}{\partial w_{\mu}^{(1)}} = x_{\mu} \quad (D1.7)$$

$$\frac{\partial \mathcal{F}(G_i)}{\partial b^{(1)}} = 1 \quad (D1.8)$$

$$\frac{\partial \mathcal{F}(G_i)}{\partial w_{\mu v}^{(0)}} = w_{\mu}^{(1)}(1 - x_{\mu}^2) G_{i,v} \quad (D1.9)$$

$$\frac{\partial \mathcal{F}(G_i)}{\partial b_{\mu}^{(0)}} = w_{\mu}^{(1)}(1 - x_{\mu}^2) \quad (D1.10)$$

Based on the Equations (3) and (C3.6), the gradient of the partial force with respect to the neural network parameters can be obtained. For instance, the derivative with respect to $w_{\mu}^{(1)}$ reads:

$$\frac{\partial}{\partial w_{\mu}^{(1)}}\left(\frac{\partial E_i}{\partial r_{ij}}\right) = \frac{\partial}{\partial w_{\mu}^{(1)}}\left(\frac{\partial E_i}{\partial G_{i,v}} \frac{\partial G_{i,v}}{\partial r_{ij}}\right) = \frac{\partial G_{i,v}}{\partial r_{ij}} \frac{\partial}{\partial w_{\mu}^{(1)}}\left(\frac{\partial E_i}{\partial G_{i,v}}\right) \quad (D1.11)$$

Since the term $\frac{\partial}{\partial w_{\mu}^{(1)}}\left(\frac{\partial E_i}{\partial G_{i,v}}\right)$ is not directly known, it can be determined by switching the order of differentiation:

$$\frac{\partial^2 E_i}{\partial w_{\mu}^{(1)} \partial G_{i,v}} = \frac{\partial^2 E_i}{\partial G_{i,v} \partial w_{\mu}^{(1)}} = \frac{\partial}{\partial G_{i,v}}\left(\frac{\partial E_i}{\partial w_{\mu}^{(1)}}\right) = \frac{\partial x_{\mu}}{\partial G_{i,v}} \quad (D1.12)$$

Substituting Equation $(D1.12)$ into $(D1.11)$:

$$\frac{\partial}{\partial w_\mu^{(1)}}\left(\frac{\partial E_i}{\partial r_{ij}}\right) = \frac{\partial G_{i,v}}{\partial r_{ij}} \frac{\partial x_\mu}{\partial G_{i,v}} \qquad (D1.13)$$

Similarly, the gradient with respect to $w_{\mu v}^{(0)}$ is given by:

$$\frac{\partial}{\partial w_{\mu v}^{(0)}}\left(\frac{\partial E_i}{\partial r_{ij}}\right) = \frac{\partial G_{i,v}}{\partial r_{ij}} \frac{\partial^2 E_i}{\partial w_{\mu v}^{(0)} \partial G_{i,v}}$$

Here, the second-order derivative takes the form:

$$\frac{\partial^2 E_i}{\partial w_{\mu v}^{(0)} \partial G_{i,v}} = \frac{\partial}{\partial G_{i,v}}\left(\frac{\partial E_i}{\partial w_{\mu v}^{(0)}}\right) = \frac{\partial}{\partial G_{i,v}}\left(w_\mu^{(1)}(1-x_\mu^2)G_{i,v'}\right)$$

$$= w_\mu^{(1)}\left(\frac{\partial(1-x_\mu^2)}{\partial G_{i,v}}G_{i,v'} + \frac{\partial G_{i,v'}}{\partial G_{i,v}}(1-x_\mu^2)\right)$$

$$= \begin{cases} w_\mu^{(1)}\left(-2w_{\mu v'}^{(0)}x_\mu(1-x_\mu^2)G_{i,v} + (1-x_\mu^2)\right), & \text{if } v = v' \\ w_\mu^{(1)}\left(-2w_{\mu v'}^{(0)}x_\mu(1-x_\mu^2)G_{i,v}\right), & \text{otherwise} \end{cases} \qquad (D1.14)$$

For the bias terms, we obtain:

$$\frac{\partial}{\partial b^{(1)}}\left(\frac{\partial E_i}{\partial r_{ij}}\right) = 0 \qquad (D1.15)$$

$$\frac{\partial}{\partial b_\mu^{(0)}}\left(\frac{\partial E_i}{\partial r_{ij}}\right) = \frac{\partial G_{i,v}}{\partial r_{ij}} \frac{\partial^2 E_i}{\partial b_\mu^{(0)} \partial G_{i,v}} \qquad (D1.16)$$

$$\frac{\partial^2 E_i}{\partial b_\mu^{(0)} \partial G_{i,v}} = w_\mu^{(1)}\left(\frac{\partial(1-x_\mu^2)}{\partial G_{i,v}}\right) = -2w_\mu^{(1)}w_{\mu v}^{(0)}x_\mu(1-x_\mu^2) \qquad (D1.17)$$

Analogously, the gradients of the virial tensor with respect to the network parameters can be derived by replacing $\frac{\partial G_{i,v}}{\partial r_{ij}}$ in the expressions above with $\frac{\partial G_i}{\partial r_{ji}} \otimes r_{ij}$, as indicated by Equations (4) and $(C3.18)$. The detailed derivation is omitted for brevity.

As derived from Equations (13)–(16), the gradients required for updating the network parameters $p\left(w_{\mu v}^{(0)}, w_\mu^{(1)}, b_\mu^{(0)}, b^{(1)}\right)$ can be obtained accordingly:

$$\frac{\partial \mathcal{L}}{\partial p} = \lambda_e \frac{\partial \mathcal{L}_e}{\partial p} + \lambda_f \frac{\partial \mathcal{L}_f}{\partial p} + \lambda_v \frac{\partial \mathcal{L}_v}{\partial p} \tag{D1.18}$$

$$\frac{\partial \mathcal{L}_e}{\partial p} = \frac{2}{N_{str}} \sum_{n=0}^{N_{str}} \left( E_i^{pre}(n) - E_i^{tar}(n) \right) \times \frac{\partial E_i}{\partial p} \tag{D1.19}$$

$$\frac{\partial \mathcal{L}_f}{\partial p} = \frac{2}{3N} \sum_{i,\alpha} \left( F_{i,\alpha}^{pre} - F_{i,\alpha}^{tar} \right) \times \frac{\partial F_i}{\partial p} \tag{D1.20}$$

$$\frac{\partial \mathcal{L}_v}{\partial p} = \frac{2}{6N_{str}} \sum_{n=0}^{N_{str}} \sum_{l=0}^{6} \left( V_{i,l}^{pre}(n) - V_{i,l}^{tar}(n) \right) \times \frac{\partial V_i}{\partial p} \tag{D1.21}$$

2. Implementation of the Neural Network

We now present an efficient CUDA device function that implements the forward and backward passes of a single-hidden-layer neural network, including: (a) Energy prediction, (b) First and second-order derivatives with respect to input descriptors, (c) Gradient propagation to network parameters.

```
1   static __device__ void apply_ann_one_layer_w2nd(
2     const int N_des,              // Number of descriptor features
3     const int N_neu,              // Number of hidden neurons
4     const float* w0,              // Weights input → hidden
5     const float* b0,              // Biases of hidden layer
6     const float* w1,              // Weights hidden → output
7     const int N,                  // Number of atoms
8     float* q,                     // Input descriptors
9     float& energy,                // Output energy
10    float* energy_derivative,     // dE/dq[n]
11    float* energy_derivative2,    // d²E/dq[n]dq[m]
12    float* ep_wb,                 // dE_wrt_wb / dq[n]
13    float* e_wb_grad)             // dE / dw0, db0, dw1, db1
14  {
15    const int offset_b0 = N_des * N_neu;
16    const int offset_w1 = offset_b0 + N_neu;
17    for (int j = 0; j < N_neu; ++j) {
18      const int j_N_des = j * N_des;
19      float w0_times_q = 0.0f;
20      for (int n = 0; n < N_des; ++n) {
21        w0_times_q += w0[j_N_des + n] * q[n];
22      }
23      const float x1 = tanh(w0_times_q + b0[j]);
```

```
24        const float tanh_der = 1.0f - x1 * x1;
25        const float tanh_der2 = -2.0f * x1 * tanh_der;    // second derivative
26        const float w1j = w1[j];
27        const float delta_1 = w1j * tanh_der;
28        energy += w1j * x1;
29        for (int n = 0; n < N_des; ++n) {
30           const int idx_w0 = j_N_des + n;
31           const float w0jn = w0[idx_w0];
32           float tmp1 = tanh_der * w0jn; // derivative of tanh w.r.t. q[n]
33           float tmp2 = w1j * tanh_der2;
34           energy_derivative[n] += w1j * tmp1;
35           // derivative of e_wb_grad[w1] w.r.t. q[n]
36           ep_wb[(offset_w1 + j) * N_des + n] = tmp1;
37           // derivative of e_wb_grad[b0] w.r.t. q[n]
38           ep_wb[(offset_b0 + j) * N_des + n] = tmp2 * w0jn;
39           const float tmp2_qn = tmp2 * q[n];
40           for (int m = 0; m < N_des; ++m) {
41              const int idx_m = j_N_des + m;
42              const float w0jm = w0[idx_m];
43              const float tmp3 = tanh_der2 * w0jn * w0jm;
44              energy_derivative2[(n * N_des + m) * N] += w1j * tmp3;
45              // derivative of e_wb_grad[w0] w.r.t. q[n]
46              ep_wb[idx_w0 * N_des + m] = tmp2_qn * w0jm;
47              ep_wb[idx_w0 * N_des + m] += (m == n) ? delta_1 : 0.0f;
48           }
49           e_wb_grad[idx_w0] += delta_1 * q[n]; // energy w.r.t. w0
50        }
51        e_wb_grad[offset_b0 + j] += delta_1; // energy w.r.t. b0
52        e_wb_grad[offset_w1 + j] += x1; // energy w.r.t. w1
53     }
54     e_wb_grad[offset_w1 + N_neu] = 1.0f; // energy w.r.t. b1
55     energy += w1[N_neu];
56 }
```

The remaining gradient computations are not listed due to their complexity. Readers interested in the implementation details are encouraged to consult the source code.

# REFERENCES

1. Batatia, I., Kovács, D. P., Simm, G. N. C., Ortner, C. & Csányi, G. MACE: Higher Order Equivariant Message Passing Neural Networks for Fast and Accurate Force Fields.

2. Behler, J. Four Generations of High-Dimensional Neural Network Potentials. *Chem. Rev.* **121**, 10037–10072 (2021).

3. Wang, Y. *et al.* On the design space between molecular mechanics and machine learning force fields. *Appl. Phys. Rev.* **12**, 021304 (2025).

4. Deng, B. *et al.* CHGNet as a pretrained universal neural network potential for charge-informed atomistic modelling. *Nat. Mach. Intell.* **5**, 1031–1041 (2023).

5. Behler, J. Perspective: Machine learning potentials for atomistic simulations. *J Chem Phys* (2024).

6. Li, K. & Ma, H. Decoding the thermal conductivity of ionic covalent organic frameworks: Optical phonons as key determinants revealed by neuroevolution potential. *Mater. Today Phys.* **54**, 101724 (2025).

7. Klarbring, J. & Walsh, A. Na Vacancy-Driven Phase Transformation and Fast Ion Conduction in W-Doped Na3SbS4 from Machine Learning Force Fields. *Chem. Mater.* (2024) doi:10.1021/acs.chemmater.4c00936.

8. Li, K., Liu, B., Zhou, J. & Sun, Z. Revealing the crystallization dynamics of Sb–Te phase change materials by large-scale simulations. *J. Mater. Chem. C* **12**, 3897–3906 (2024).

9. Yan, Z. & Zhu, Y. Impact of Lithium Nonstoichiometry on Ionic Diffusion in

Tetragonal Garnet-Type Li7La3Zr2O12. *Chem. Mater.* **36**, 11551–11557 (2024).

10. Fransson, E., Wiktor, J. & Erhart, P. Phase Transitions in Inorganic Halide Perovskites from Machine-Learned Potentials. *J. Phys. Chem. C* **127**, 13773–13781 (2023).

11. Ying, P. *et al.* Atomistic insights into the mechanical anisotropy and fragility of monolayer fullerene networks using quantum mechanical calculations and machine-learning molecular dynamics simulations. *Extreme Mech. Lett.* **58**, 101929 (2023).

12. Wang, B. *et al.* Thermal Transport of GeTe/$Sb_2Te_3$ Superlattice by Large-Scale Molecular Dynamics with Machine-Learned Potential. *J. Phys. Chem. C* **129**, 6386–6396 (2025).

13. Eckhoff, M. & Behler, J. High-dimensional neural network potentials for magnetic systems using spin-dependent atom-centered symmetry functions. *Npj Comput. Mater.* **7**, 1–11 (2021).

14. Shapeev, A. V. Moment Tensor Potentials: A Class of Systematically Improvable Interatomic Potentials. *Multiscale Model. Simul.* **14**, 1153–1173 (2016).

15. Thompson, A. P. Spectral neighbor analysis method for automated generation of quantum-accurate interatomic potentials. *J. Comput. Phys.* (2015).

16. Bartók, A. P., Payne, M. C., Kondor, R. & Csányi, G. Gaussian Approximation Potentials: The Accuracy of Quantum Mechanics, without the Electrons. *Phys. Rev. Lett.* **104**, 136403 (2010).

17. Fan, Z. Neuroevolution machine learning potentials: Combining high accuracy and low cost in atomistic simulations and application to heat transport. *Phys. Rev. B*

(2021).

18. Schaul, T., Glasmachers, T. & Schmidhuber, J. High dimensions and heavy tails for natural evolution strategies. in *Proceedings of the 13th annual conference on Genetic and evolutionary computation* 845–852 (Association for Computing Machinery, New York, NY, USA, 2011). doi:10.1145/2001576.2001692.

19. Fan, Z. *et al.* GPUMD: A package for constructing accurate machine-learned potentials and performing highly efficient atomistic simulations. *J. Chem. Phys.* **157**, 114801 (2022).

20. Lenc, K., Elsen, E., Schaul, T. & Simonyan, K. Non-Differentiable Supervised Learning with Evolution Strategies and Hybrid Methods. Preprint at https://doi.org/10.48550/arXiv.1906.03139 (2019).

21. Kingma, D. P. & Ba, J. Adam: A Method for Stochastic Optimization. Preprint at https://doi.org/10.48550/arXiv.1412.6980 (2017).

22. Loshchilov, I. & Hutter, F. Decoupled Weight Decay Regularization. Preprint at https://doi.org/10.48550/arXiv.1711.05101 (2019).

23. Kiyani, E., Shukla, K., Urbán, J. F., Darbon, J. & Karniadakis, G. E. Which Optimizer Works Best for Physics-Informed Neural Networks and Kolmogorov-Arnold Networks? Preprint at https://doi.org/10.48550/arXiv.2501.16371 (2025).

24. Frye, C. G. *et al.* Critical Point-Finding Methods Reveal Gradient-Flat Regions of Deep Network Losses. *Neural Comput.* **33**, 1469–1497 (2021).

25. Wierstra, D. *et al.* Natural Evolution Strategies. *J. Mach. Learn. Res.* **15**, 949–980 (2014).

26. Wen, T., Zhang, L., Wang, H., E, W. & Srolovitz, D. J. Deep potentials for materials science. *Mater. Futur.* **1**, 022601 (2022).

27. Podryabinkin, E. V. Active learning of linearly parametrized interatomic potentials. *Comput. Mater. Sci.* (2017).

28. Behler, J. Atom-centered symmetry functions for constructing high-dimensional neural network potentials. *J. Chem. Phys.* **134**, 074106 (2011).

29. Fan, Z. *et al.* Force and heat current formulas for many-body potentials in molecular dynamics simulations with applications to thermal conductivity calculations. *Phys. Rev. B* **92**, 094301 (2015).

30. Kirkpatrick, S., Gelatt, C. D. & Vecchi, M. P. Optimization by Simulated Annealing. *Science* **220**, 671–680 (1983).

31. Goyal, P. *et al.* Accurate, Large Minibatch SGD: Training ImageNet in 1 Hour. *arXiv.org* https://arxiv.org/abs/1706.02677v2.

# Supporting Information

Efficient GPU-Accelerated Training of a Neuroevolution Potential with Analytical Gradients

Hongfu Huang,[a] Junhao Peng,[b] Kaiqi Li,[c] Jian Zhou,[d] and Zhimei Sun[*d]

[a] **School of Integrated Circuit Science and Engineering, Beihang University, Beijing, 100191, China.**

[b] **Guangdong Provincial Key Laboratory of Sensing Physics and System Integration Applications, School of Physics and Optoelectronic Engineering, Guangdong University of Technology, Guangzhou, Guangdong 510006, China.**

[c] **National Key Laboratory of Spintronics, Hangzhou International Innovation Institute, Beihang University, Hangzhou 311115, China.**

[d] School of Materials Science and Engineering, Beihang University, Beijing 100191, China.

[*] **Corresponding author: zmsun@buaa.edu.cn**

## I. Training based different datasets

To evaluate the general applicability of gradient-based NEP training, we extended our comparison between SNES and Adam optimizers across a range of representative datasets. For each case, we report training performance in terms of RMSE for energy, force, and virial components as a function of wall-clock time and training epoch.

A. Silicon

The Si dataset, adapted from Ref. [1], includes a wide range of configurations: bulk crystal structures, sp-bonded and sp²-bonded structures, β-Sn structures, dimers, liquid phases, amorphous phases, decohesion configurations, diamond structures with surfaces or vacancies, and several other defective structures.

Figure S1 compares training convergence using SNES and Adam on this dataset.

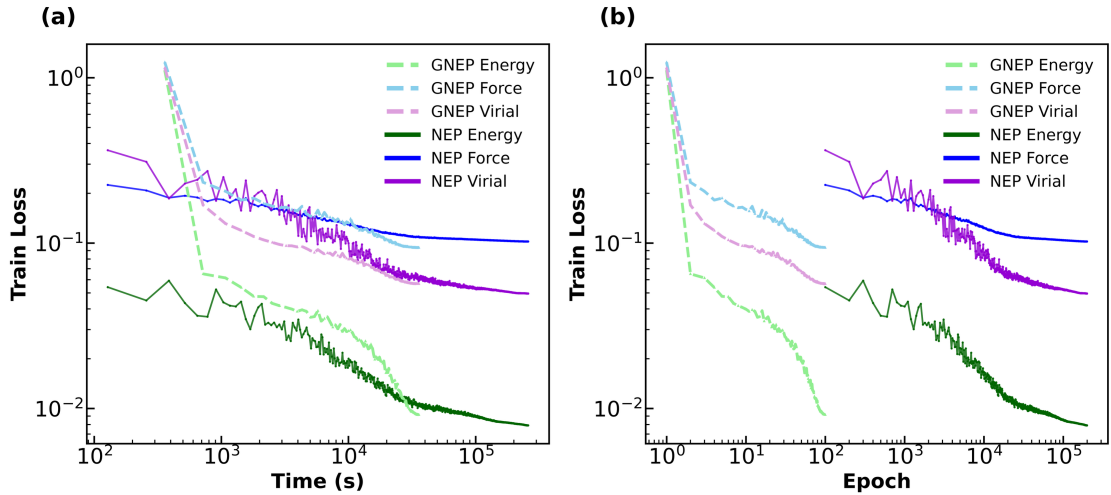

**Figure S1** Training RMSE for energy, force and virial predictions on the Si dataset, comparing SNES (solid lines) and Adam optimizers (dash–dot lines). (a) Training loss vs. wall-clock time. (b) Training loss vs. the number of epochs.

B. SiO$_2$

This part uses the SiO$_2$ dataset described in Ref. [2], which includes a diverse set of configurations such as 1832 crystal ones, 761 amorphous or liquid ones, and 16 dimer or cluster ones.

Figure S2 presents the corresponding training performance comparison.

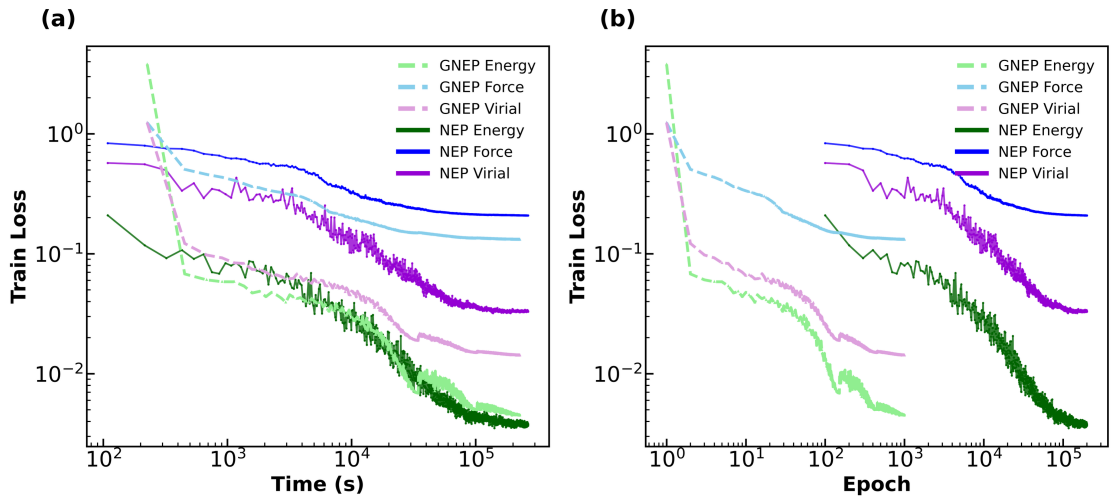

**Figure S2** Training RMSE for energy, force and virial predictions on the SiO$_2$ dataset, comparing SNES (solid lines) and Adam optimizers (dash–dot lines). (a) Training loss vs. wall-clock time. (b) Training loss vs. the number of epochs.

C. GeTe/Sb$_2$Te$_3$

We trained the NEP model using a dataset of 1346 GeTe/Sb$_2$Te$_3$ superlattice configurations adapted from Ref. [3]. The dataset comprises various superlattice structures with GeTe: Sb$_2$Te$_3$ layer ratios (1:1, 1:2, 1:3, 1:4, 2:2, 2:6, and 2:8), as well as configurations with antisite defects and randomly applied strain or angular distortion. A total of 821 configurations were used for training and 525 for testing.

Training results can be seen in Figure S3.

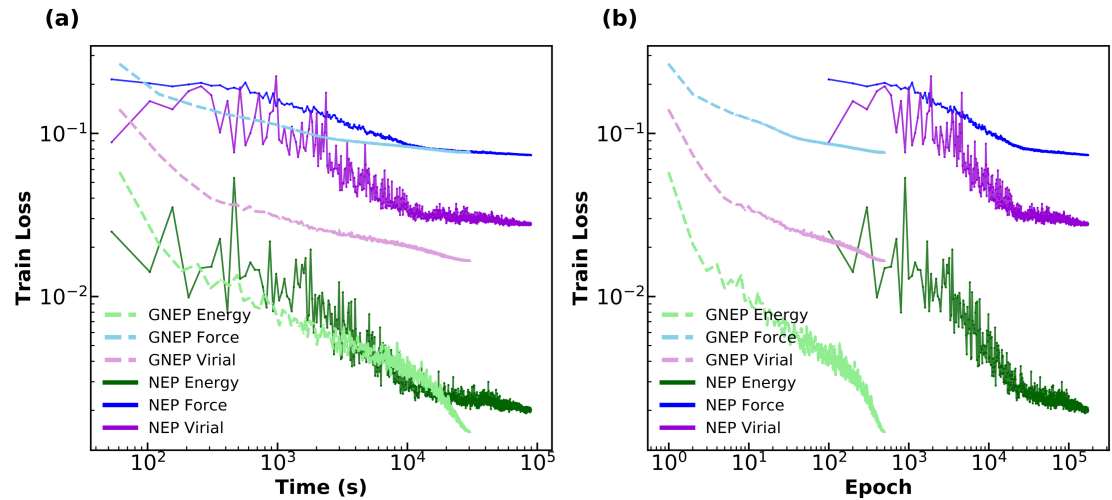

**Figure S3** Training RMSE for energy, force and virial predictions on the GeTe/Sb$_2$Te$_3$ superlattice dataset, comparing SNES (solid lines) and Adam optimizers (dash–dot lines). (a) Training loss vs. wall-clock time. (b) Training loss vs. the number of epochs.

D. PdCuNiP

The dataset, adapted from Ref. [4], comprises 10,525 configurations of Pd-Cu-Ni-P alloys, including unary to quaternary compositions. It spans FCC, BCC, and HCP lattice types, and includes structures with surfaces, vacancies, and thermal perturbations across different temperatures. The training and test sets contain 9615 and 910 configurations, respectively.

Training results are shown in Figure S4.

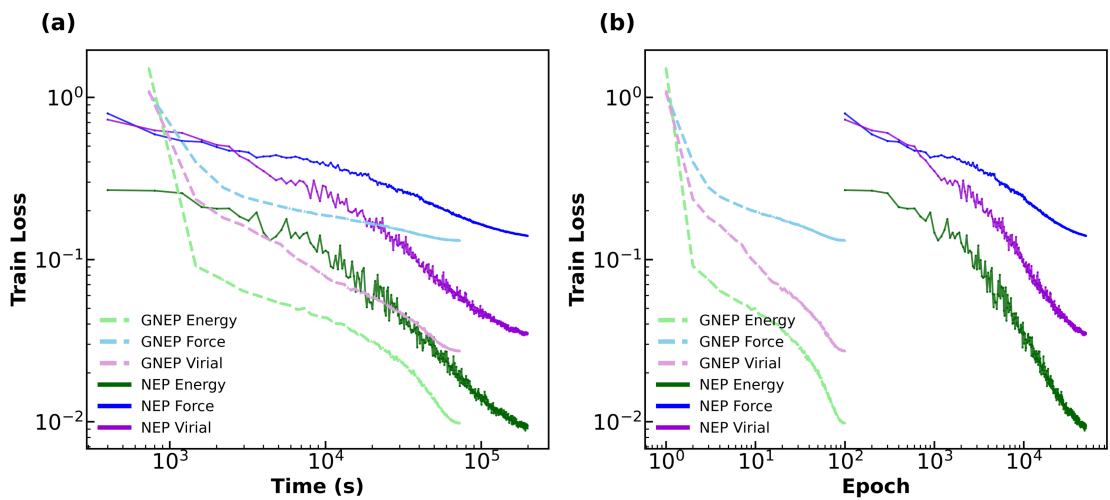

**Figure S4** Training RMSE for energy, force and virial predictions on the PdCuNiP dataset, comparing SNES (solid lines) and Adam optimizers (dash–dot lines). (a) Training loss vs. wall-clock time. (b) Training loss vs. the number of epochs.

E. ICOF-10n-Li/Na (n = 1, 2, and 3)

This dataset, based on Ref. [5], includes ICOF-10n-Li/Na (n = 1, 2, and 3) structures, and corresponding configurations relaxed within a temperature range from 250 K to 450 K during MD simulations.

Training results are shown in Figure S5.

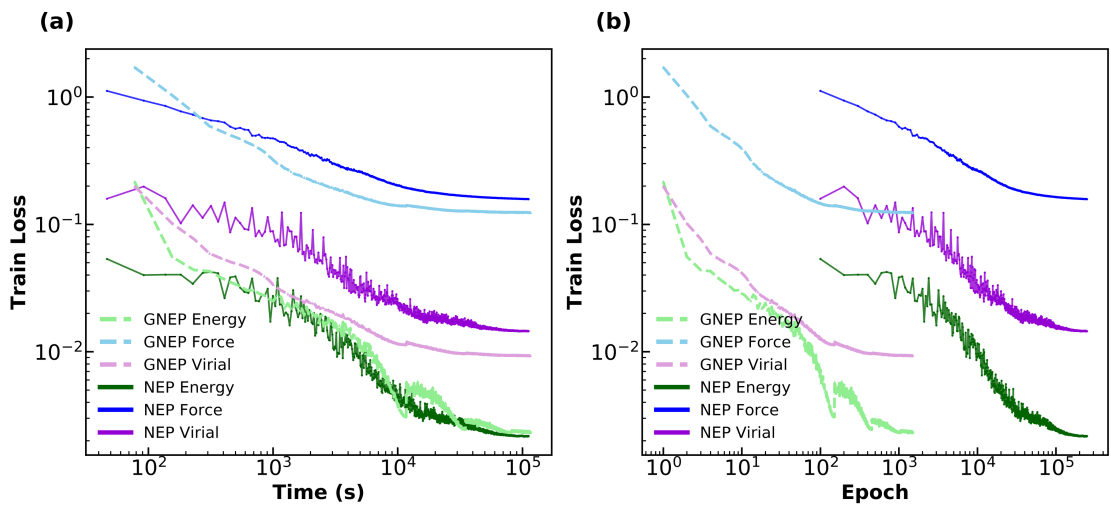

**Figure S5** Training RMSE for energy, force and virial predictions on the ICOF-10n-Li/Na (n = 1, 2, and 3) dataset, comparing SNES (solid lines) and Adam optimizers (dash–dot lines). (a) Training loss vs. wall-clock time. (b) Training loss vs. the number of epochs.

# References

1.	Bartók, A. P., Kermode, J., Bernstein, N. & Csányi, G. Machine Learning a General-Purpose Interatomic Potential for Silicon. *Phys. Rev. X* **8**, 041048 (2018).

2.	Liang, T. *et al.* Mechanisms of temperature-dependent thermal transport in amorphous silica from machine-learning molecular dynamics. *Phys. Rev. B* **108**, 184203 (2023).

3.	Wang, B. *et al.* Thermal Transport of GeTe/Sb$_2$Te$_3$ Superlattice by Large-Scale Molecular Dynamics with Machine-Learned Potential. *J. Phys. Chem. C* **129**, 6386–6396 (2025).

4.	Zhao, R. *et al.* Development of a neuroevolution machine learning potential of Pd-Cu-Ni-P alloys. *Mater. Des.* **231**, 112012 (2023).

5.	Li, K. & Ma, H. Decoding the thermal conductivity of ionic covalent organic frameworks: Optical phonons as key determinants revealed by neuroevolution potential. *Mater. Today Phys.* **54**, 101724 (2025).